\documentclass[aps,eqsecnum,pra,twocolumn,longbibliography]{revtex4}
\usepackage{epsfig,latexsym,amsmath,amssymb,amscd,graphicx,amsbsy,amsfonts,bbm}
\usepackage{graphicx,hyperref}
\begin{document}
\title{On `orbital' and `spin' angular momentum of light in 
classical and quantum theories -- a general framework}
\author{Arvind}
\email{arvind@iisermohali.ac.in}
\affiliation{Department of Physical Sciences,
Indian
Institute of Science Education \&
Research (IISER) Mohali, Sector 81 SAS Nagar,
Manauli PO 140306 Punjab India.}
\author{S. Chaturvedi}
\email{subhash@iiserbhopal.ac.in}
\affiliation{Department of Physics,
Indian
Institute of Science Education \&
Research (IISER) Bhopal, Bhopal Bypass Road, Bhauri, Bhopal
462066 India}
\author{N. Mukunda}
\email{nmukunda@gmail.com}
\affiliation{
INSA Distinguished Professor,
Indian Academy of Sciences,
C V Raman Avenue, Sadashivanagar,
Bangalore 560080 India}
\begin{abstract}
We develop a
general framework to analyze the
two important and much discussed questions concerning (a)
`orbital' and `spin' angular momentum carried by light and
(b) the paraxial approximation of the free Maxwell system
both in the classical as well as quantum domains.  After
formulating the classical free Maxwell system in the
transverse gauge  in terms of complex analytical signals we
derive expressions for the constants of motion associated
with its Poincar\'{e} symmetry.  In particular, we show that
the constant of motion corresponding to the total angular
momentum ${\bf J}$ naturally splits into an `orbital' part
${\bf L}$ and a `spin' part ${\bf S}$ each of which is a
constant of motion in its own right. A noteworthy feature of
the formulation so developed is the emergence of a complex
Hilbert space $\mathcal{M}$ associated with the free Maxwell
system which in turn provides a natural link between the
descriptions of radiation fields at the classical and
quantum levels.  We then proceed to discuss quantization of
the free Maxwell system and construct the operators
generating the Poincar\'{e} group in the quantum context and
analyze their algebraic properties and find that while the
quantum counterparts $\hat{{\bf L}}$ and $\hat{{\bf S}}$ of
${\bf L}$ and ${\bf S}$ go over into bona fide observables,
they fail to satisfy the angular momentum algebra precluding
the possibility of their interpretation as `orbital' and
`spin' operators at the classical level.  On the other hand
$\hat{{\bf J}}=\hat{{\bf L}}+ \hat{{\bf S}}$ does satisfy
the angular momentum algebra and together with $\hat{{\bf
S}}$ generates the group $E(3)$.  We then present an
analysis of single photon states, paraxial
quantization both in the scalar as well as vector cases,
single photon states in the paraxial regime.
All along a close connection is
maintained with the Hilbert space $\mathcal{M}$ that arises
in the classical  context thereby providing a bridge between
classical and quantum descriptions of radiation fields.  The
present approach provides strong motivation for making the
paraxial approximation after, and not before, field
quantization has been carried out.

\end{abstract} 
\maketitle
\section{Introduction}
There has been much interest recently in the theoretical
understanding and experimental exploitation of the angular
momentum properties of light in both classical and quantum
domains~\cite{babiker-book}. It has been shown that the
total angular momentum of the (free) electromagnetic field,
a well-defined expression which is conserved on account of
Poincar\'{e} invariance, can be meaningfully separated into
`orbital' and `spin' parts, each of which is conserved by
itself~\cite{wolf-book}.  (For brevity the former will be
called the OAM -- orbital angular momentum -- of light).
Through experiments it has been suggested that specific
effects attributable to the OAM of light can be identified.
Schemes to filter OAM states have been
proposed~\cite{apt2,jha}
and proposals for beam splitters for the OAM states aimed at
achieving quantum information processing have also been
developed~\cite{apt5}. There have also been several
discussions on the cloning of OAM states~\cite{apt6} and
network-based OAM~\cite{apt7}.  The quantum states with OAM
have been proposed to be useful for implementation of
quantum gates~\cite{apt4} and for  universal quantum
computation~\cite{apt1,apt3}.  Quantum key distribution
schemes based on OAM have also been proposed\cite{apt8,apt9}
and the possibility of creating entangled photons has also
been discussed~\cite{apt10}.  Entanglement of OAM states has
been demonstrated experimentally~\cite{ape1}. Experiments
have also been performed on implementing  quantum random
walks~\cite{ape2} and quantum cryptographic schemes using
quantum states of light with OAM~\cite{ape4}. Squeezing of
states with OAM has been experimentally
demonstrated~\cite{ape5} and various correlation experiments
have also been carried out~\cite{ape6,ape7}.  Single photon
in modes with OAM have been prepared~\cite{ape8} and
Laguerre-Gaussian laser modes with OAM have also been
generated~\cite{ape9}.

The properties of OAM of light for the practically
important paraxial beams are important as laser beam modes
are of this nature.  Another aspect is to see
whether anything new can be learnt about the quantum
mechanics of single photons, both in terms of possible
physical observables and in the properties of particular
quantum states. 
A comprehensive analysis is contained in the work of Calvo
et.  al.~\cite{calvo-pra-06}, in which a systematic
approximation scheme for paraxial beams has been proposed.
In addition the procedure to be adopted for field
quantization to arrive at a physically reasonable concept of
`paraxial photons' has been addressed by Deutsch et.
al.~\cite{deutsch-pra-91}.  Here it may be recalled that
single photons with fixed helicity are described by specific
mass zero unitary irreducible representations (UIR's) of the
Poincar\'{e} group not including the parity
operation~\cite{wigner-am-39}. A photon with right (or left)
circular polarization remains so under all proper rotations
as well as under all pure Lorentz transformations. Examining
these representations in detail, Newton and
Wigner~\cite{newton-rmp-49} found in a classic analysis that
no position operator with reasonable physical properties can
be defined for photons. This result can be expected to have
important consequences for the existence of OAM as a
meaningful and measurable property for single photons, and
therefore also for the quantized field.

The purpose of the present work is to analyze carefully the
properties and interconnections of the different concepts
mentioned above -- OAM for classical light; the specific
algebraic properties needed for a vectorial dynamical
variable to be an `angular momentum'; related properties at
the level of single photon states; and the details of the
paraxial limit for light in relation to quantization. The
paraxial regime for light beams is by its very nature
defined in a qualitative and approximate way. Therefore the
relationship between it and the quantization process needs
careful analysis. In effect, the concept of `paraxial
photons' has to be arrived at in a physically reasonable
manner. In addition, there is the pervasive effect of
transversality in all matters relating to electromagnetism
-- the fields and vector potential; the quantization
process; photon wave functions, and even the definition of
the paraxial regime.

The contents of the paper are organized as follows. In
Section II we discuss the Poincar\'{e} covariance of the
Maxwell equations in free space and derive, in the
transverse radiation gauge, expressions for the associated
constants of motion in terms of the real fields ${\bf
E}({\bf x},t)$, the electric field and ${\bf A}({\bf x},t)$,
the vector potential. With eventual passage to a description
of quantized electromagnetic fields in mind we move from the
real ${\bf A}$ and ${\bf E}$ fields to their complex
analytic signals ${\bf A}^{(+)}$ and ${\bf E}^{(+)}$ and
re-express the ten constants of motion  in terms of ${\bf
A}^{(+)}({\bf x},t)$ and ${\bf E}^{(+)}({\bf x},t)$ and also
in terms of the complex transverse ${\bf k}$-space
amplitudes ${\bf v}({\bf k})$ associated with ${\bf
A}^{(+)}({\bf x},t)$. In particular, we note that the
constant of motion, ${\bf J}$, the total angular momentum,
can be regarded as a sum of `orbital' and `spin' parts ${\bf
L}$ and ${\bf S}$ each of which is also a constant of motion
in its own right. The structure of the expressions for the
constants of motion obtained thus prompts us to introduce
the notion of a Hilbert space $\mathcal{M}$ with ${\bf
A}^{(+)}({\bf x},t)$ or equivalently their complex
transverse ${\bf k}$-space amplitudes ${\bf v}({\bf k})$ as
its elements.  In Section III we quantize the free space
Maxwell field following the canonical route by turning the
${\bf k}$-space amplitudes ${\bf v}({\bf k})$ into operators
$\hat{{\bf a}}({\bf k})$ and the fundamental Poisson
brackets into canonical commutation relations. This then
sets the stage for introducing the notion of  Hilbert spaces
$\mathcal{H}_n$ with $n=0,1,2,\cdots $ spanned respectively
of no photon or the vacuum state, 1 photon states, 2 photon
states ....and hence to the $\mathcal{H}$ associated with
quantized electromagnetic field obtained by taking their
direct sum. The analytic signals ${\bf A}^{(+)}$ and ${\bf
E}^{(+)}$ now become operators $\hat{{\bf A}}^{(+)}$ and
$\hat{{\bf E}}^{(+)}$ and the classical constants of motions
go over into hermitian operators on ${\cal H}$.  With this
done we compute all possible commutation relations between
the components of the quantum counterparts   $\hat{{\bf J}},
\hat{{\bf L}}$ and $\hat{{\bf S}}$ of ${\bf J}, {\bf L}$ and
${\bf S}$, and find that while the components of $\hat{{\bf
J}}$ obey the angular momentum algebra, those of $\hat{{\bf
L}}$ and $\hat{{\bf S}}$ fail to do so. Further while the
components of $\hat{{\bf S}}$  commute with each other,
those of $\hat{{\bf L}}$ and $\hat{{\bf S}}$ do not all
commute. This circumstance calls into question decomposition
of the total angular momentum operator into an `orbital'
angular momentum part $\hat{{\bf L}}$  and a `spin' angular
momentum part $\hat{{\bf S}}$ as has already been noted in
the literature.

Next we define single photon states in the one photon
Hilbert space $\mathcal{H}_1$ by associating
creation-annihilation operators $\hat{a}({\bf v})^\dagger$,
$\hat{a}({\bf v})$ with each ${\bf v}({\bf k})$ or
equivalently with each ${\bf A}^{(+)}({\bf x},t)$ in
$\mathcal{M}$. We then detail  their  properties  and
actions of the operators corresponding to the Poincar\'{e}
algebra and of the infinitesimal unitary transformations
they generate on these states.

In Section IV, we turn our attention to the paraxial
approximation in  classical scalar and vector optics with a
view to giving a meaningful definition of `quantization in
the paraxial regime'. We show, in contrast to several
works in  the literature, that a better procedure is to impose
paraxial conditions on ${\bf v}({\bf k})$ which appear in
the definition of the one-photon states as is discussed in Section
III.
We conclude this
Section with a discussion of 
Laguerre--Gaussian(L-G) modes studied extensively
in scalar wave optics anticipating their extension to the
vector case in the following Section. In Section V we
construct the eigenfunctions of $\hat{J}_3$ in the single
photon subspace both in the ${\bf k}$ space as well as in
the physical space and examine their behaviors in the
paraxial regime. We also extend the scalar L-G modes
discussed in the previous Section to the vector case.
Section VI contains a summary of this work and our
concluding remarks. 
\section{Classical free Maxwell system, real fields and
analytic signals, constants of motion} \label{maxwell} In
this Section, as background, we recall basic properties of
the classical Maxwell equations, and settle matters of
notation. Certain features not often discussed explicitly
are mentioned.

The equations of motion (EOM) for real electric and magnetic
fields in free space are
\begin{eqnarray}
\dfrac{\partial}
{\partial t}{\bf E}({\bf x},t) &=& c
\mbox{\bf{$\nabla$}}\wedge{\bf B}({\bf x},t),\nonumber \\
\dfrac{\partial}{\partial t}{\bf B}({\bf x},t) &=&
- c \mbox{\bf{$\nabla$}}\wedge{\bf E}({\bf x},t),
\label{maxwell1}
\end{eqnarray}
supplemented by the constraints
\begin{equation}
 \mbox{\bf{$\nabla$}}\cdot{\bf E}({\bf x},t)=
\mbox{\bf{$\nabla$}}\cdot {\bf B}({\bf x},t)=0
\label{maxwell2}
\end{equation}
which are preserved in time. In special relativistic
notation ($x^0=ct; g_{00}=-1; \mu, \nu=0, 1, 2, 3$), ${\bf
E}$ and ${\bf B}$ are components of a real antisymmetric
second rank tensor field $F_{\mu\nu}(x)=-F_{\nu\mu}(x),
(x=(ct, {\bf x}))$:
\begin{equation}
 E_j(x)=F_{j0}(x),\,\, B_j(x)=\dfrac{1}{2}
\epsilon_{jlm}F_{lm}(x), \,\, j,l,m=1,2,3.
\label{maxwell_covariant1}
\end{equation}
Then eqs.~(\ref{maxwell1}, \ref{maxwell2}) are
\begin{equation}
\partial^{\mu}F_{\mu\nu}(x) =0,~~~~
\partial_{\lambda}F_{\mu\nu}(x)+
\partial_{\mu}F_{\nu\lambda}(x)+\partial_{\nu}F_{\lambda\mu}(x)=0.
\label{maxwell_covariant2}
\end{equation}
These are manifestly covariant under the action of
inhomogeneous Lorentz, i.e., Poincar\'{e}, transformations.
A general element $(\Lambda, a)$ of the Poincar\'{e} group
${\cal P}$ acts on space-time coordinates $x=(x^{\mu})$ by
\begin{equation}
 x^\mu\rightarrow x^{\prime\mu}=\Lambda^{\mu}_{~\nu}~ x^\nu+ a^{\mu}.
\label{maxwell_covariant3}
\end{equation}
Here $\Lambda=(\Lambda^\mu_{~\nu})\in O(3,1)$ is a real
$4\times 4$ matrix of a homogeneous Lorentz transformation
in which we allow parity but omit time reversal, and $a^\mu$
is a space time translation. The composition law in ${\cal
P}$ is
\begin{equation}
(\Lambda^\prime,a^\prime)(\Lambda,a)=(\Lambda^\prime\Lambda,
a^\prime+\Lambda^\prime a),
\label{maxwell_covariant4}
\end{equation}
while $F_{\mu\nu}$ transforms as
\begin{equation}
F^{\prime}_{\mu\nu}(x^\prime)=
\Lambda_{\mu}^{~\rho}\Lambda_{\nu}^{~\sigma}F_{\rho\sigma}(x).
\label{maxwell_covariant5}
\end{equation}
This action by ${\cal P}$ preserves
eqs.~(\ref{maxwell_covariant2}).

We hereafter work in the transverse radiation gauge, to
begin with using real fields ${\bf E}({\bf x}, t), {\bf
A}({\bf x}, t)$. The EOM, constraints and definition of
${\bf B}$ are:
\begin{subequations}
\begin{eqnarray}
 \dfrac{\partial}{\partial t}{\bf A}({\bf x},t)&=&
-c{\bf E}({\bf x},t),
 \nonumber \\
\dfrac{\partial}{\partial t}{\bf E}({\bf x},t)&=&
-c\nabla^2{\bf A}({\bf x},t);
\\
\mbox{\bf{$\nabla$}}\cdot{\bf E}({\bf x},t)&=&
\mbox{\bf{$\nabla$}}\cdot{\bf A}({\bf x},t)=0;
 \\
 {\bf B}({\bf x},t)&=&\mbox{\bf{$\nabla$}}\wedge{\bf A}({\bf x},t).
\end{eqnarray}\label{vector_potential}
\end{subequations}
The behaviors of ${\bf E}$ and ${\bf A}$ under spatial
rotations and space-time translations are immediate. Under
an infinitesimal pure Lorentz transformation we first
transform ${\bf A}(x)$ as though it is a four-vector (with
$A_0=0$), and then perform an infinitesimal gauge
transformation to restore transversality:
\begin{eqnarray}
 x^{\prime 0}&\simeq& x^{0}-\dfrac{v_j}{c}x_j,\quad
x^{\prime}_j\simeq x_{j}-\dfrac{v_j}{c} x^0,\quad
|{\bf v}|<< c:\nonumber\\
{\bf E}^\prime(x^\prime)&\simeq&
{\bf E}(x)+\dfrac{1}{c}{\bf v}\wedge{\bf B}(x),\nonumber\\
{\bf A}^\prime(x^\prime)&\simeq&{\bf
A}(x)+\frac{1}{c}\mbox{\bf{$\nabla$}}\left(\frac{1}{\nabla^2}{\bf
v}\cdot {\bf E}\right)(x)\nonumber\\
&=&{\bf A}(x)-\frac{1}{4\pi c}\mbox{\bf{$\nabla$}}
\int d^3y \dfrac{{\bf v}\cdot{\bf E}({\bf y},t)}{|{\bf x}-{\bf y}|}.
\label{low_velocity}
\end{eqnarray}
The basic constants of motion (COM), ten in number, associated with
solutions to eqs.~(\ref{maxwell1},~\ref{vector_potential})
are the energy-momentum four-vector $P^\mu$ and
four-dimensional angular momentum  second rank antisymmetric
tensor $J_{\mu\nu}$. These are
expressed in terms of an energy density ${\mathcal E}({\bf
x}, t)$ and a momentum density
$\mbox{\boldmath$\cal{P}$}({\bf x}, t)$:
\begin{eqnarray}
{\mathcal E}({\bf x}, t)&=&\frac{1}{8\pi}({\bf E}({\bf x}, t)\cdot{\bf
E}({\bf x}, t)+{\bf B}({\bf x}, t)
\cdot{\bf B}({\bf x}, t)),\nonumber\\
\mbox{\boldmath$\cal{P}$}({\bf x}, t)&=&\frac{1}{4\pi c}{\bf
E}({\bf x}, t)\wedge{\bf B}({\bf x}, t),
\label{energy_momenturm}
\end{eqnarray}
(the Poynting vector being $c^2\mbox{\boldmath$\cal{P}$}$) as
\begin{eqnarray}
P^{0}&=&\int d^3x~ {\mathcal E}({\bf x},t),~~~{\bf P}=\int
d^3x ~\mbox{\boldmath$\cal{P}$}({\bf x},t);\nonumber\\
{\bf J}&=& (\text{J}_{23}, \text{J}_{31}, \text{J}_{12})
=\int d^3x ~{\bf x}\wedge \mbox{\boldmath$\cal{P}$}({\bf
x},t);\nonumber\\
{\bf K}&=& (\text{J}_{10}, \text{J}_{20}, \text{J}_{30}) =
\dfrac{1}{c}\int d^3x ~{\bf x}~{\cal E} ({\bf x},t)-ct\text{{\bf P}}.
\nonumber \\
\label{p_j_k}
\end{eqnarray}
In terms of ${\bf E}$ and ${\bf A}$, dropping surface terms at spatial infinity we have:
\begin{eqnarray}
&&P^{0}=\dfrac{1}{8\pi}\int d^3x~ ( {\bf E}({\bf x},t)
\cdot {\bf E}({\bf x},t)- {\bf A}({\bf x},t)\cdot \nabla^2{\bf A}({\bf x},t)), \nonumber\\
&&{\bf P}=\dfrac{1}{4\pi c}\int d^3x ~{E}_m({\bf x},t)
\mbox{\boldmath$\nabla$}{A}_m({\bf x},t) ;\nonumber\\
&&J_j=\dfrac{1}{4\pi c}\int d^3x ~{E}_m({\bf
x},t)(\delta_{mn} ({\bf x}\wedge
\mbox{\boldmath$\nabla$})_j+\epsilon_{jmn}){A}_n({\bf
x},t),\nonumber\\
&&K_j(t) =\dfrac{1}{c}\int d^3x ~x_j~{\cal E} ({\bf x},t)\nonumber\\
&&=
\dfrac{1}{8\pi c}\int d^3x~ x_j ({\bf E}({\bf x},t)\cdot
{\bf E}({\bf x},t)- {\bf A}({\bf x},t)\cdot \nabla^2{\bf A}({\bf x},t)).
\nonumber \\
\label{momentum_and_angular}
\end{eqnarray}
Equations (\ref{vector_potential}) lead to the wave equation
for ${\bf A}(x)$. Omitting possible evanescent waves, the
general solution can be 
written as
\begin{eqnarray}
&&{\bf A}({\bf x},t)=\dfrac{c}{2\pi}
\int_{\mathbb{R}^3}\dfrac{d^3k}{\sqrt{\omega}}(e^{ik\cdot x}{\bf v}({\bf k})
+\text{c.c}),\nonumber \\
&&{\bf k}\cdot{\bf v}({\bf k})=0, \,
k^0=|{\bf k}|= \omega/c.
\label{vpot_fourier}
\end{eqnarray}
Thus it involves one complex transverse vector amplitude
${\bf v}({\bf k})$ on wave vector space. The real electric
field is
\begin{equation}
{\bf E}({\bf x},t)=\dfrac{i}{2\pi}\int  d^3k ~\sqrt{\omega}~(e^{ik\cdot x}{\bf v}({\bf k})
-\text{c.c}).
\label{efield_fourier}
\end{equation}
In terms of ${\bf v}({\bf k})$, the COM's and ${\bf K}(t)$ are:
\begin{eqnarray}
(P^{0}, {\bf P})&=& \int d^3k(ck^0, {\bf k})~ {\bf v}({\bf k})^\ast\cdot {\bf
v}({\bf k}),\nonumber \\
J_j&=&\int d^3k ~{v}_m({\bf k})^\ast(-i\delta_{mn} ({\bf
k}\wedge
\tilde{\mbox{\boldmath$\nabla$}})_j-i\epsilon_{jmn}){v}_n({\bf
k}) ,\nonumber\\
K_j(t) &=&\dfrac{i}{c}\int d^3k ~\sqrt{\omega}~{\bf v} ({\bf
k})^\ast \cdot \tilde{\partial}_j~(\sqrt{\omega}~{\bf v}
({\bf k}))+ct\ P_j ,
\nonumber\\
\tilde{\partial}_j&\equiv& \frac{\partial}{\partial k_j}.
\label{epjk_fourier}
\end{eqnarray}

At this point, ${\bf v}({\bf k})$ is usually expanded in
terms of a basis of two polarization vectors
$\mbox{\boldmath$\epsilon$}_\alpha({\bf k}), 
\alpha= 1, 2$, obeying
\begin{eqnarray}
& {\bf k}\cdot\mbox{\boldmath$\epsilon$}_\alpha({\bf k})
=0,\quad
 \mbox{\boldmath$\epsilon$}_\alpha({\bf k})^\ast
\cdot\mbox{\boldmath$\epsilon$}_\beta({\bf
k})=\delta_{\alpha\beta},& \nonumber \\
& \mbox{\boldmath$\epsilon$}_1({\bf k})
\wedge\mbox{\boldmath$\epsilon$}_2({\bf k})=\hat{{\bf k}},&
\label{transverse}
\end{eqnarray}
as
\begin{equation}
 {\bf v}({\bf k})=\sum_{\alpha=1}^{2}
\mbox{\boldmath$\epsilon$}_\alpha({\bf k}) v_\alpha ({\bf k}).
\label{vector_v}
\end{equation}
In general we can allow
$\mbox{\boldmath$\epsilon$}_\alpha({\bf k})$ to depend on
${\bf k}$, though sometimes we may assume dependence on 
$\hat{\bf k}$ alone. This will be clear from the context.

However, there are subtleties associated with choosing such
polarization  bases in a globally smooth way as $\hat{\bf
k}$ varies over all possible directions, so we take this up
later~\cite{nityananda-ap-14, mukunda-josaa-14,
arvind-pla-17}

We now move from real ${\bf A}, {\bf E}$ to their complex
analytic signals ${\bf A}^{(+)}, {\bf E}^{(+)}$.  For the
definition and properties of analytic signals see,
for instance~\cite{born-book,wolf-book}.
This gives the classical theory a 
form similar to quantum expressions. The analytic signals
are the positive frequency parts of the real fields:
\begin{eqnarray}
{\bf E}&=&{\bf E}^{(+)}+ {\bf E}^{(-)}=2\text{Re}~{\bf
E}^{(+)},\nonumber \\
{\bf A}&=&{\bf A}^{(+)}+ {\bf A}^{(-)}=2\text{Re}~{\bf A}^{(+)}.
\label{positive}
\end{eqnarray}
While eqs.~(\ref{vector_potential}) are obeyed after the
replacements ${\bf A}, {\bf E}, {\bf B}\rightarrow {\bf
A}^{(+)}, {\bf E}^{(+)}, {\bf B}^{(+)}$, 
we also have first order EOM in time which are nonlocal in
space:
\begin{eqnarray}
i\dfrac{\partial}{\partial t}{\bf A}^{(+)}({\bf x},t)
&=&(\hat{\omega}{\bf A}^{(+)})({\bf x},t), \nonumber \\
i\dfrac{\partial}{\partial t}{\bf E}^{(+)}({\bf
x},t)&=&(\hat{\omega}{\bf E}^{(+)})({\bf x},t),\nonumber\\
 \hat{\omega}&=& c (-\mbox{\boldmath$\nabla$}^2)^{1/2}
\label{wave_eqns}
\end{eqnarray}
The general
solution~(\ref{vpot_fourier},\ref{efield_fourier}) 
becomes
\begin{eqnarray}
{\bf A}^{(+)}({\bf x},t)&=&\dfrac{c}{2\pi}
\int\dfrac{d^3k}{\sqrt{\omega}}e^{ik\cdot x}{\bf v}({\bf k}),
\nonumber \\
{\bf E}^{(+)}({\bf x},t)&=&
\dfrac{i}{2\pi}\int d^3k~\sqrt{\omega}
e^{ik\cdot x}{\bf v}({\bf k}).
\label{AE_fromvk}
\end{eqnarray}
The COM's (\ref{momentum_and_angular}) can all be written in
terms of analytic signals. We need to use the hermiticity of
$\hat{\omega}$ in the 
sense that for suitable complex $\psi({\bf x}), \phi({\bf
x})$ we have
\begin{equation}
\int d^3x~\phi({\bf x})^\ast(\hat{\omega}\psi)({\bf x})=\int
d^3x~((\hat{\omega}\phi)({\bf x}))^\ast\psi({\bf x}).
\label{exchange}
\end{equation}
Then after some algebra we find:
\begin{eqnarray}
 {P}^{0}&=&\dfrac{1}{2\pi}\int d^3x~
{\bf E}^{(+)}(x)^\ast\cdot \partial^0{\bf A}^{(+)}(x),
\nonumber \\
P_j&=&\dfrac{1}{2\pi c}\int d^3x ~{\bf E}^{(+)}(x)^\ast
\cdot \partial_j{\bf A}^{(+)}(x),\nonumber\\
J_j&=&\dfrac{1}{2\pi c}\int d^3x ~{ E}_m^{(+)}(x)^\ast
\left(\delta_{mn}({\bf x}\wedge
\mbox{\boldmath$\nabla$})_j \right. \nonumber\\
&&\left. \quad\quad\quad+\epsilon_{jmn}\right){ A}_n^{(+)}(x) ,\nonumber\\
K_j(t) &=&
\dfrac{i}{2\pi c^2}\int d^3x~ {\bf E}^{(+)}(x)^\ast\cdot
(x_j\hat{\omega}{\bf A}^{(+)})(x).\nonumber\\
\label{Ps_Js_and_K}
\end{eqnarray}
Thus eqs.~(\ref{momentum_and_angular},\ref{Ps_Js_and_K},\ref{epjk_fourier})
give ${P}^\mu, {\bf J}, {\bf K}(t)$ in terms of real fields, 
complex analytic signals, and ${\bf k}$-space amplitudes respectively.

In the real formulation, ${\bf E}({\bf x}, 0)$ and ${\bf
A}({\bf x}, 0)$ can be chosen independently, then  eqs.~(\ref{vector_potential}) 
determine ${\bf E}(x), {\bf A}(x)$ for all $t$. In the
complex formulation, only ${\bf A}^{(+)}({\bf x}, 0)$ can be
chosen independently. 
Then both ${\bf A}^{(+)}(x)$ and ${\bf
E}^{(+)}(x)=\partial^0{\bf
A}^{(+)}(x)=\frac{i}{c}(\hat{\omega}{\bf A}^{(+)})(x)$ are
determined by eq.~(\ref{wave_eqns}).

The structure of the total angular momentum ${\bf J}$ in eq.~(\ref{Ps_Js_and_K}) has led to the suggestion that we regard
it as the 
sum of `orbital' and `spin' parts ${\bf L}$ and ${\bf S}$,
each real, defined as~\cite{enk-epl-94}
\begin{eqnarray}
L_j&=&\dfrac{1}{2\pi c}\int d^3x ~{ E}_m^{(+)}(x)^\ast ({\bf
x}\wedge \mbox{\boldmath$\nabla$})_j{A}_m^{(+)}(x)
,\nonumber\\
S_j&=&\dfrac{1}{2\pi c}\int d^3x\ \epsilon_{jmn}
~{E}_m^{(+)}(x)^\ast{A}_n^{(+)}(x).
\label{Ls_Ss}
\end{eqnarray}
From the first order EOM (\ref{wave_eqns}), the hermiticity
(\ref{exchange}) of $\hat{\omega}$, and the 
commutativity of $\hat{\omega}$ and ${\bf x}\wedge
\mbox{\boldmath$\nabla$}$, we see that both ${\bf L}$ and
${\bf S}$ are COM's. 
Since they are unambiguously defined expressions, they are
legitimate classical dynamical variables.

The forms of the expressions (\ref{Ps_Js_and_K}) suggest
that we introduce a complex Hilbert space at the classical
level as follows. 
Guided by eq.~(\ref{exchange}) we initially  define a
natural looking inner product among complex (transverse)
vector fields 
${\bf V}({\bf x}), {\bf V}'({\bf x}), \ldots$ on $\mathbb{R}^3$ as
\begin{equation}
 \left({\bf V}^\prime(\cdot)~,~{\bf V}(\cdot)\right)_0=\int
d^3x~{\bf V}^\prime({\bf x})^\ast\cdot{\bf V}({\bf x}).
\label{vdotv}
\end{equation}
Then we find
\begin{subequations}
\begin{eqnarray}
\left({\bf V}^\prime(\cdot)~,~\hat{\omega}{\bf
V}(\cdot)\right)_0&=& \left(\hat{\omega}{\bf
V}^\prime(\cdot)~,~{\bf V}(\cdot)\right)_0;
\label{vdotv_2}\\
\left({\bf V}^\prime(\cdot),\mbox{\boldmath$\alpha$}
\cdot {\bf x}\wedge \mbox{\boldmath$\nabla$}{\bf
V}(\cdot)\right)_0&=&
 - \left( \mbox{\boldmath$\alpha$}\cdot {\bf x}\wedge
\mbox{\boldmath$\nabla$}{\bf V}^\prime(\cdot),{\bf
V}(\cdot)\right)_0 ,\nonumber \\
&&\mbox{\boldmath$\alpha$}\in
 \mathbb{R}^3.
\label{vcrossv}
\end{eqnarray}\label{vcross}
\end{subequations}
From given analytic signal solutions ${\bf A}^{(+)}(x), {\bf
E}^{(+)}(x)$ to the Maxwell equations we can form three
inner products 
$({\bf A}^{(+)}(\cdot, t), {\bf A}^{(+)}(\cdot, t))_0, ({\bf
E}^{(+)}(\cdot, t), {\bf A}^{(+)}(\cdot, t))_0$ and 
$({\bf E}^{(+)}(\cdot, t), {\bf E}^{(+)}(\cdot, t))_0$. From
eq.~(\ref{vdotv_2}) all of them are time independent.
However we now show that 
only the second one is also Lorentz invariant.

The separations (\ref{positive}) of ${\bf A}(x), {\bf E}(x)$
into positive and negative frequency parts are Lorentz
invariant. Therefore from 
eqs.~(\ref{low_velocity}) we can read off the changes in
functional forms of ${\bf A}^{(+)}(x), {\bf E}^{(+)}(x)$
under an infinitesimal 
Lorentz transformation:
\begin{eqnarray}
\delta{\bf A}^{(+)}(x)&\equiv&{\bf A}^{(+)\prime}(x)-{\bf
A}^{(+)}(x)\nonumber\\
&=&t{\bf v}\cdot \mbox{\boldmath$\nabla$}{\bf
A}^{(+)}(x)-\dfrac{{\bf v}\cdot{\bf x}}{c}~{\bf
E}^{(+)}(x)\nonumber\\
&+&\dfrac{1}{c}\mbox{\boldmath$\nabla$}\left(\dfrac{1}{\mbox{\boldmath$\nabla$}^2}{\bf
v}\cdot {\bf E}^{(+)}\right)(x), \nonumber\\
\delta{\bf E}^{(+)}(x)&\equiv&{\bf E}^{(+)\prime}(x)-{\bf E}^{(+)}(x)\nonumber\\
&=&t{\bf v}\cdot \mbox{\boldmath$\nabla$}{\bf E}^{(+)}(x)-\dfrac{{\bf v}\cdot{\bf x}}{c}
\mbox{\boldmath$\nabla$}^2{\bf A}^{(+)}(x)\nonumber\\
&+&\dfrac{1}{c}{\bf v}\wedge {\bf B}^{(+)}(x).
\label{deltaA_deltaE}
\end{eqnarray}
Then, using transversality and dropping surface terms, we find:
\begin{eqnarray}
&&\delta \left({\bf E}^{(+)}(\cdot,t)~,~{\bf
A}^{(+)}(\cdot,t)\right)_0\nonumber\\
&&=\left(\delta{\bf E}^{(+)}(\cdot,t)~,~{\bf
A}^{(+)}(\cdot,t)\right)_0+ \left({\bf E}^{(+)}(\cdot,t)~,
~\delta{\bf A}^{(+)}(\cdot,t)\right)_0\nonumber\\
&&=\dfrac{1}{c}\int d^3x ~{\bf v}\wedge {\bf B}^{(+)}(x)^\ast
\cdot{\bf A}^{(+)}(x) \nonumber \\
&&\quad +\dfrac{i}{c^2}\int d^3x ~{\bf E}^{(+)}(x)^\ast
\cdot ([\hat{\omega},{\bf v}\cdot {\bf x}]{\bf A}^{(+)})(x)\nonumber\\
&&=\dfrac{1}{c}\int d^3x ~{\bf v}\wedge
( \mbox{\boldmath$\nabla$}\wedge {\bf A}^{(+)}(x))^\ast \cdot{\bf A}^{(+)}(x)
\nonumber \\
&&\quad
-i\int d^3x~ {\bf E}^{(+)}(x)^\ast \cdot\left({\bf v}\cdot
\mbox{\boldmath$\nabla$}\dfrac{1}{\hat{\omega}}{\bf
A}^{(+)}\right)(x)\nonumber\\
&&=\dfrac{1}{c}\int d^3x~ (\mbox{\boldmath$\nabla$}
{\bf v}\cdot {\bf A}^{(+)}(x)-{\bf v}
\cdot \mbox{\boldmath$\nabla$} {\bf A}^{(+)}(x))^\ast\cdot{\bf A}^{(+)}(x)
\nonumber \\
&&\quad
-\dfrac{1}{c}\int d^3x~ (\hat{\omega}{\bf A}^{(+)})(x)^\ast
\cdot\left({\bf v}.
\mbox{\boldmath$\nabla$}\dfrac{1}{\hat{\omega}}
{\bf A}^{(+)}\right)(x)\nonumber\\
&&=0.
\label{deltaEdotA}
\end{eqnarray}
Here we used the operator relations
\begin{equation}
[\hat{\omega}^2,{\bf x}]=-2c^2\mbox{\boldmath$\nabla$},
\quad
[\hat{\omega},{\bf x}]=-c^2\dfrac{\mbox{\boldmath$\nabla$}}{\hat{\omega}}.
\label{omega_comm}
\end{equation}
Therefore $({\bf E}^{(+)}(\cdot, t), {\bf A}^{(+)}(\cdot,
t))_0$ is both time independent and Lorentz invariant.

We point out that for real ${\bf E}, {\bf A}$ the expression
$\int d^3x\ {\bf E}(x)\cdot {\bf A}(x)$ is neither Lorentz
invariant nor time independent.
Both these facts can be traced to the freedom to choose
${\bf E}({\bf x}, 0)$ and ${\bf A}({\bf x}, 0)$
independently as initial data. 
This emphasizes the advantages of using analytic signals.

Based on these considerations we define a Lorentz invariant
and time independent squared norm for any analytic signal as
\begin{eqnarray}
&&\left({\bf A}^{(+)}(\cdot,t)~,~{\bf
A}^{(+)}(\cdot,t)\right)\nonumber \\
&&\quad
=\dfrac{i}{2\pi c} \left({\bf E}^{(+)}(\cdot,t)~,~{\bf
A}^{(+)}(\cdot,t)\right)_0\nonumber\\
&&\quad=\dfrac{1}{2\pi c^2}\left({\bf
A}^{(+)}(\cdot,t)~,~\hat{\omega}{\bf
A}^{(+)}(\cdot,t)\right)_0\nonumber\\
&&\quad=\dfrac{1}{2\pi c^2}\int d^3x\ {\bf A}^{(+)}(x)^\ast
\cdot(\hat{\omega}{\bf A}^{(+)})(x)\ge 0.\nonumber\\
\label{AplusAplus}
\end{eqnarray}
Since
\begin{equation}
({\bf A}^{(+)}(\cdot,t)~,~\hat{\omega}{\bf
A}^{(+)}(\cdot,t))_0=2\pi c^2\int d^3k\ {\bf v}({\bf
k})^\ast \cdot {\bf v}({\bf k}),
\label{Aplus1}
\end{equation}
we are led to a complex Hilbert space ${\cal M}$ at the
classical level.
This Hilbert space at the classical level has been used
in~\cite{mukunda-pramana-86}
It has also been used in~\cite{deutsch-pra-91} where however,
its Lorentz invariance has not been discussed or proved.
\begin{eqnarray}
{\cal M}\!&=&\!\left\{{\bf A}^{(+)}(x)|{\|{\bf
A}^{(+)}\|^2}\!=\!\left({\bf A}^{(+)}
(\cdot,t)~,~{\bf A}^{(+)}(\cdot,t)\right) <
\infty\right\}\nonumber\\
&=&\bigg\{{\bf v}({\bf k}) ~|~
{\bf k}\cdot {\bf v}({\bf k})=0, \|{\bf
v}\|^2\nonumber\\&=&\left. \int d^3k~
{\bf v}({\bf k})^\ast \cdot {\bf v}({\bf k}) < \infty\right\}.
\nonumber \\
\label{setM}
\end{eqnarray}
We can think of either ${\bf A}^{(+)}(x)$ or ${\bf v}({\bf
k})$ connected by eq.~(\ref{AE_fromvk}) as specifying an
element of ${\cal M}$. 
The usefulness of this construction will emerge as we
proceed to the quantized Maxwell field and the quantum
mechanics of single photons. 
Both of eqs.~(\ref{vcross}) remain valid with the new inner product:
\begin{eqnarray}
&&\left({\bf A}^{(+)\prime}(\cdot,t)~,~
\hat{\omega} {\bf A}^{(+)}(\cdot,t)\right)
\nonumber \\
&&\quad=  \left( \hat{\omega} {\bf
A}^{(+)\prime}(\cdot,t)~,~{\bf
A}^{(+)}(\cdot,t)\right),\nonumber\\
&&\left({\bf A}^{(+)\prime}(\cdot,t)~,~
\mbox{\boldmath$\alpha$}\cdot {\bf x}
\wedge\mbox{\boldmath$\nabla$} {\bf A}^{(+)}(\cdot,t)\right)
\nonumber \\
&&\quad =-  \left( \mbox{\boldmath$\alpha$}\cdot {\bf
x}\wedge\mbox{\boldmath$\nabla$} {\bf
A}^{(+)\prime}(\cdot,t)~,~{\bf A}^{(+)}(\cdot,t)
\right).
\label{AplusdeltaAplus}
\end{eqnarray}
\section{The quantized radiation field, operator CoM's,
quantum mechanics of single photons}
In this Section we briefly recall the canonical quantization
of the Maxwell field, and study the properties of the basic
operator COM's and the 
counterparts of ${\bf L}$ and ${\bf S}$ of
eq.~(\ref{Ls_Ss}). We then look at some features of single
photon states and operator actions on them.
\subsection*{Quantization and basic operator COM's}
Since the free Maxwell equations (\ref{vector_potential})
are linear, we can first construct their general solution
and then perform quantization. 
In the analytic signal formulation, the general solution
(\ref{AE_fromvk}) involves the complex transverse vector
amplitude ${\bf v}({\bf k})$. 
Consistent with transversality the Hamiltonian form of the
classical theory contains the basic Poisson Bracket (PB)
relations
\begin{eqnarray}
&&\{v_j({\bf k}), v_l({\bf k}')^\ast\}=
-i\left(\delta_{jl}-\dfrac{k_jk_l}{|{\bf
k}|^2}\right)\delta^{(3)}({\bf k}-{\bf
k}^\prime),\nonumber\\
&& \{{\bf v}, {\bf v}\}= \{{\bf v}^\ast, {\bf v}^\ast\}=0.
\label{PB}
\end{eqnarray}
Using the Dirac prescription `quantum commutators' $\sim
i\hbar$  `classical PB's', we arrive at the canonical
commutation relations (CCR) in a 
convenient form:
\begin{eqnarray}
&&{\bf v}({\bf k})\rightarrow \sqrt{\hbar}~\hat{{\bf a}}({\bf k}),\quad {\bf v}({\bf k})^\ast\rightarrow \sqrt{\hbar}~\hat{{\bf a}}({\bf k})^\dagger:\nonumber\\
&& [\hat{a}_j({\bf k}), \hat{a}_l({\bf k}^\prime)^\dagger] = \left(\delta_{jl}-\dfrac{k_jk_l}{|{\bf k}|^2}\right)\delta^{(3)}({\bf k}-{\bf k}^\prime),\nonumber\\&& [{\bf \hat{a}}, {\bf \hat{a}}]= [{\bf \hat{a}}^\dagger, {\bf \hat{a}}^\dagger]=0,\nonumber\\
&&{\bf k}\cdot \hat{a}({\bf k})={\bf k}\cdot \hat{a}({\bf k})^\dagger=0.
\label{CCR}
\end{eqnarray}
As with eqs.~(\ref{transverse}, \ref{vector_v}) the
transversality conditions can be accommodated by expanding 
$\hat{\bf a}({\bf k}), \hat{\bf a}({\bf k})^\dagger$ in terms of $ \mbox{\boldmath$\epsilon$}_\alpha({\bf k})$:
\begin{eqnarray}
&&\hat{{\bf a}}({\bf k})=\sum_{\alpha=1}^{2} \mbox{\boldmath$\epsilon$}_\alpha({\bf k})\hat{{a}}_\alpha({\bf k}),\nonumber\\
&&\hat{{\bf a}}({\bf k})^\dagger=\sum_{\alpha=1}^{2} \mbox{\boldmath$\epsilon$}_\alpha({\bf k})^\ast\hat{{a}}_\alpha({\bf k})^\dagger;\nonumber\\
&&[\hat{a}_\alpha({\bf k}), \hat{a}_\beta({\bf k}^\prime)^\dagger] = \delta_{\alpha\beta}\delta^{(3)}({\bf k}-{\bf k}^\prime),\nonumber\\
&& [\hat{a},\hat{a}]=[\hat{a}^\dagger,\hat{a}^\dagger]=0.
\label{poln_basis}
\end{eqnarray}
We mention again that there are subtleties involved in making globally smooth choices of such polarization bases which have been properly 
analyzed only recently~\cite{
nityananda-ap-14, mukunda-josaa-14, arvind-pla-17}. It has
of course been known that if one limits oneself to, say,
linear polarizations, no globally smooth choices exist. 
Once one allows general polarization states with complex $ \mbox{\boldmath$\epsilon$}_\alpha({\bf k})$'s, global smoothness can be achieved. 
We introduce chosen polarization bases later when needed.

The space-time dependent operator analytic signals and hermitian field operators are:
\begin{subequations}
\begin{eqnarray}
&& \hat{{\bf A}}^{(+)}(x)=\dfrac{c}{2\pi}\sqrt{\hbar}\int \dfrac{d^3k}{\sqrt{\omega}}~e^{ik\cdot x}\hat{{\bf a}}({\bf k}),\nonumber\\
&& \hat{ {\bf E}}^{(+)}(x)=\dfrac{i}{2\pi}\sqrt{\hbar}\int d^3k~ \sqrt{\omega}e^{ik\cdot x}\hat{{\bf a}}({\bf k});\\
&&\hat{{\bf A}}(x)=\hat{{\bf A}}^{(+)}(x)+\hat{{\bf A}}^{(-)}(x),\nonumber\\
&&\hat{{\bf E}}(x)=\hat{{\bf E}}^{(+)}(x)+\hat{{\bf E}}^{(-)}(x).
\end{eqnarray}  \label{field_operators}
\end{subequations}

The Hilbert space ${\cal H}$ on which all the above operators act and realize their CR's is the direct sum of subspaces labelled by the total 
photon number $n=0, 1, 2, \ldots$:
\begin{equation}
{\cal H}=\sum\limits_{\oplus n=0}^\infty{\cal H}_n.
\label{H_S}
\end{equation}
${\cal H}_0$ is spanned by the vacuum state $|0\rangle$ with no photons:
\begin{eqnarray}
&&{\cal H}_0=\{\lambda|0\rangle,~\lambda\in\mathbb{C}\},\nonumber\\
 &&\hat{{\bf a}}({\bf k})|0\rangle=0,~~\langle 0|0\rangle=1.
 \label{vacuum}
\end{eqnarray}
The subspace ${\cal H}_1$ consists of all single photon states, obtained by applying all possible (normalisable) `linear combinations' 
of $\hat{{\bf a}}({\bf k})^\dagger$ to $|0\rangle$:
\begin{equation}
  {\cal H}_1=\text{Sp}\left\{\hat{a}_j({\bf
k})^\dagger|0\rangle~|~{\bf k}
\in\mathbb{R}^3,~j=1,2,3\right\}.
  \label{one_photon}
\end{equation}
Action by $\hat{{\bf a}}({\bf k})^\dagger$'s on ${\cal H}_1$ leads to the two-photon subspace ${\cal H}_2$, and so on.

The basic classical COM's written in eq.~(\ref{Ps_Js_and_K})
in terms of analytic signals give immediately the hermitian
quantum operator 
COM's in normal ordered form. They generate the unitary operators representing elements of the Poincar\'{e} group in the quantum theory. 
We list all of them including the pure Lorentz transformation generators:
\begin{eqnarray}
\hat{P}_{0}&=&\dfrac{i}{2\pi c}\int d^3x~ \hat{{\bf
E}}^{(+)}(x)^\dagger\cdot (\hat{\omega}\hat{{\bf
A}}^{(+)})(x)\nonumber\\
&=&\int d^3k ~\hbar\omega~
\hat{{\bf a}}({\bf k})^\dagger \cdot \hat{{\bf a}}({\bf k});\nonumber\\
\hat{P}_j&=&\dfrac{1}{2\pi c}\int d^3x ~\hat{{\bf
E}}^{(+)}(x)^\dagger \cdot \partial_j\hat{{\bf
A}}^{(+)}(x)\nonumber\\
&=& \int d^3k ~\hbar k_j ~\hat{{\bf a}}({\bf k})^\dagger\cdot \hat{{\bf a}}({\bf k});\nonumber\\
\hat{J}_j&=&\dfrac{1}{2\pi c}\int d^3x ~\hat{E}_m^{(+)}(x)^\dagger (\delta_{mn} ({\bf x}\wedge \mbox{\boldmath$\nabla$})_j+\epsilon_{jmn})
\hat{A}_n^{(+)}(x) \nonumber\\
&=&-i\hbar \int d^3k ~\hat{a}_m({\bf k})^\dagger
(\delta_{mn} ({\bf k}\wedge
\tilde{\mbox{\boldmath$\nabla$})}_j+\epsilon_{jmn})
~\hat{a}_n({\bf k});\nonumber\\
\hat{K}_j(t) &=&
\dfrac{i}{2\pi c^2}\int d^3x~ \hat{{\bf
E}}^{(+)}(x)^\dagger\cdot (x_j\hat{\omega}\hat{{\bf
A}}^{(+)})(x)\nonumber\\
&=&ct \hat{\text{P}}_j +\dfrac{i\hbar}{c}\int
d^3k~\sqrt{\omega}\hat{a}_m({\bf
k})^\dagger\tilde{\partial}_j(\sqrt{\omega}\hat{a}_m({\bf
k})).
\label{operator_COM}
\end{eqnarray}
The two hermitian COM's $\hat{\mathbf J}$ and $\hat{\mathbf K}(0)$, as is well known, realize the Lie algebra of the homogeneous Lorentz group SO(3, 1):
\begin{eqnarray}
 &&[\hat{J}_j, \hat{J}_l]=
 i\hbar~\epsilon_{jlm}~\hat{J}_m, ~~[\hat{J}_j,\hat{K}_l(0)]=i\hbar~\epsilon_{jlm}~\hat{K}_m(0),\nonumber\\
 &&[\hat{K}_j(0), \hat{K}_l(0)]= -i\hbar~\epsilon_{jlm}~\hat{J}_m.
 \label{SO_31}
\end{eqnarray}

Next we consider the quantum operator analogues of the
proposed definitions, eqs.~(\ref{Ls_Ss}), of `orbital' and
`spin' angular momenta 
${\bf L, S}$ for the free Maxwell field. From
eqs.~(\ref{operator_COM}) let us define the operators
\begin{eqnarray}
&&\hat{L}_j=-i\hbar\int d^3k~\hat{a}_m({\bf k})^\dagger({\bf
k}\wedge \tilde{\mbox{\boldmath$\nabla$}})_j\hat{a}_m({\bf
k}),\nonumber\\
&&\hat{S}_j=-i\hbar\int d^3k~\hat{a}_m({\bf k})^\dagger
\epsilon_{jmn}\hat{a}_n({\bf k}),
 \label{OAM}
\end{eqnarray}
so
\begin{equation}
 \hat{J}_j=\hat{L}_j+\hat{S}_j.
 \label{Total_AM}
\end{equation}
Each of $\hat{L}_j$ and $\hat{S}_j$ is well-defined and
hermitian (as well as a COM). Therefore each of them is a
physically observable dynamical 
variable representing a corresponding property of the
quantum field. However, whether they may be regarded as
angular momenta in the quantum 
mechanical sense depends on their CR's. The analyses of
$\hat{L}_j$ and $\hat{S}_j$ follow similar lines. We find
after some algebra:
\begin{equation}
[\hat{L}_j, \hat{L}_l]=i\hbar\ \epsilon_{jlm}(\hat{L}_m-\hat{S}_m).
\label{L_L}
\end{equation}
While the right hand side is antihermitian, the presence of
the second term shows that the $\hat{L}_j$ are not a quantum
mechanical angular momentum 
triplet. This is so despite each $\hat{L}_j$ being hermitian
and a COM. Similarly the CR's among the $\hat{S}_j$ can be
computed and are
 \begin{equation}
[\hat{S}_j, \hat{S}_l]=0.
\label{S_S}
 \end{equation}
 Therefore the $\hat{S}_j$ too do not form a quantum mechanical angular momentum. These facts concerning $\hat{L}_j$ and $\hat{S}_j$ have been noted 
 earlier~\cite{enk-epl-94}. They mean that a priori we cannot draw any conclusions about the eigenvalues of each $\hat{L}_j$ and each 
 $\hat{S}_j$, or of $\hat{\bf L}\cdot \hat{\bf L}$ and $\hat{\bf S}\cdot \hat{\bf S}$. In fact, unlike with a true angular momentum, 
 all the components $\hat{S}_j$ can be simultaneously diagonalised.

 To complete the picture, we need the commutators $[\hat{S}_j, \hat{L}_l]$ which turn out to be
\begin{equation}
[\hat{S}_j, \hat{L}_l]=i\hbar\ \epsilon_{jlm}~\hat{S}_m.
\label{S_L}
\end{equation}
We infer: (i) since these commutators are nonzero, $\hat{\bf L}$ and $\hat{\bf S}$ are not kinematically independent; (ii) by combining 
eqs.~(\ref{L_L}, \ref{S_S}, \ref{S_L}) we obtain
\begin{equation}
[\hat{J}_j, \hat{L}_l\ \ \mbox{or}\ \ \hat{S}_l]=i\hbar\ \epsilon_{jlm}(\hat{L}_m\ \ \mbox{or}\ \hat{S}_m).
\label{Vectors}
\end{equation}
Therefore both $\hat{\bf L}$ and $\hat{\bf S}$ transform as vectors under spatial SO(3) rotations generated by $\hat{\bf J}$.

It is worth noting that eqs.~(\ref{Vectors}) mean that the
following would have been equivalent: $\hat{\bf L}$ obeys
the angular momentum CR's; $\hat{\bf S}$ obeys these CR's;
$\hat{\bf S}$ and $\hat{\bf L}$ commute with one another.
These are the properties characteristic of orbital and spin
angular momenta for massive particles.  Of course, none of
these is true.  On the other hand, we see that $\hat{\bf J}$
and $\hat{\bf S}$ obey the CR's corresponding to the
Euclidean group E(3), though each of them has the dimensions
of action. 
\subsection*{Single photon states -- properties, operator actions}
The one-photon subspace ${\cal H}_1$ is defined in
eq.~(\ref{one_photon}). There is a natural and direct
correspondence between the classical Hilbert 
space ${\cal M}$ of eq.~(\ref{setM}) and ${\cal H}_1$. For
each ${\bf v}(\cdot)\in {\cal M}$ with classical analytic
signal ${\bf A}^{(+)}(x)$, 
we define a corresponding creation -- annihilation operator
pair $\hat{a}({\bf v})^\dagger, \hat{a}({\bf v})$ for
photons `in the state 
${\bf v}(\cdot)$'.
Such operators have been introduced earlier in
reference~\cite{mukunda-pramana-86} as well as in
reference~\cite{deutsch-pra-91}.
\begin{eqnarray}
&&{\bf v}({\bf k})\in {\cal M},~~{\bf A}^{(+)}({\bf
x},t)=\dfrac{c}{2\pi}\int\dfrac{d^3k}{\sqrt{\omega}}~e^{ik\cdot
x}{\bf v}({\bf k})\rightarrow\nonumber\\
&& {\hat a}({\bf v})=\dfrac{1}{\sqrt{\hbar}}\int {d^3k}~
{\bf v}({\bf k})^\ast \cdot \hat{{\bf a}}({\bf k}), ~~~
{\hat a}({\bf v})^\dagger\nonumber\\
&&=\dfrac{1}{\sqrt{\hbar}}\int {d^3k}~ {\bf v}({\bf k}) \cdot \hat{{\bf a}}({\bf k})^\dagger;~~~~~~~~~~~~~~~~~~~~~~~~~~(i)\nonumber\\
&&[{\hat a}({\bf v}),{\hat a}({\bf v}^\prime)^\dagger]=\dfrac{({\bf v},{\bf v}^\prime)}{\hbar}\mathbbm{1};
~~~~~~~~~~~~~~~~~~~~~~~~~~~~(ii)\nonumber\\
&&|{\bf v}\rangle={\hat a}({\bf v})^\dagger|0\rangle \in {\cal H}_1~~~,~~~\langle {\bf v}^\prime|{\bf v}\rangle=
\dfrac{({\bf v}^\prime,{\bf v})}{\hbar}~~~~~~(iii)\nonumber\\\label{single_photon}
\end{eqnarray}
The connection to the classical analytic signal is through
\begin{equation}
 \langle 0| \hat{{\bf A}}^{(+)}(x)|{\bf v}\rangle= {\bf A}^{(+)}(x).
 \label{anal_signal}
\end{equation}
In this sense, every classical ${\bf v}({\bf k})$ and associated ${\bf A}^{(+)}(x)$ are the momentum space and physical space wave functions 
respectively for a single photon in the state $|{\bf v}\rangle$. After normalization this state is
\begin{equation}
 {\sqrt{\hbar}}\dfrac{~|{\bf v}\rangle}{\|{\bf v}\|} = \dfrac{1}{\|{\bf v}\|}\int d^3k~ {\bf v}({\bf k})\cdot \hat{{\bf a}}({\bf k})^\dagger|0\rangle,
 \label{single_photon4}
\end{equation}
so ${\bf v}({\bf k})^\ast\cdot {\bf v}({\bf k})/\|{\bf v}\|^2$ is the ${\bf k}$-space probability density distribution. 
On the other hand, ${\bf A}^{(+)}(x)$ is the photon wave function in a formal sense, as ${\bf x}$ does not represent photon position.

Now we consider the actions of the operators $\hat{\bf J}, \hat{\bf K}(0), \hat{\bf L}\ \mbox{and}\ \hat{\bf S}$, and of the infinitesimal 
unitary transformations generated by them, on single photon states. Each of these operators conserves total photon number, so it leaves 
invariant each subspace ${\cal H}_n$ in eq.~(\ref{H_S}). The restrictions of these operators to ${\cal H}_1$ (or to any ${\cal H}_n$) will 
therefore also obey the CR's (\ref{SO_31}, \ref{L_L}, \ref{S_S}, \ref{S_L}, \ref{Vectors}). For simplicity we will use the same symbols for 
these restrictions as for the complete operators, as the meaning will be clear from the context. We take up $\hat{\bf J}, \hat{\bf K}(0)$ first.

From eqs.~(\ref{operator_COM}), using the CCR's (\ref{CCR}), we obtain the CR's among $\hat{\bf J}_j, \hat{\bf K}_j(0)$ on the one hand, and 
$\hat{a}_l({\bf k})^\dagger$ on the other. For $\hat{J}_j$ we get the set of results:
\begin{subequations}
\begin{eqnarray}
 &&[\hat{J}_j, \hat{a}_l({\bf k})^\dagger] =i\hbar(\delta_{lm}({\bf k}\wedge\tilde{\mbox{\boldmath$\nabla$}})_j+\epsilon_{jlm})\hat{a}_m({\bf k})^\dagger;\nonumber\\ \label{rotation1}\\
 &&|\mbox{\boldmath$\alpha$}|<<1:~~~~~\left(1-\dfrac{i}{\hbar}\mbox{\boldmath$\alpha$}\cdot\hat{{\bf J}}\right)|{\bf v}\rangle\simeq
 |{\bf v}+\delta{\bf v}\rangle,\nonumber\\
 &&~~~~~~~~\delta{\bf v}({\bf k})=-\mbox{\boldmath$\alpha$}\cdot{\bf k}\wedge\tilde{\mbox{\boldmath$\nabla$}}\ {\bf v}({\bf k})+
 \mbox{\boldmath$\alpha$}\wedge
 {\bf v}({\bf k});\nonumber\\\label{rotation2}\\
&& \delta({\bf v}({\bf k})^\ast\cdot{\bf v}({\bf k}))=
 \tilde{\mbox{\boldmath$\nabla$}}\cdot( -\mbox{\boldmath$\alpha$}\wedge {\bf k}\ {\bf v}({\bf k})^\ast\cdot{\bf v}({\bf k})).\nonumber\\\label{rotation3}
 \end{eqnarray}\label{rotation}
 \end{subequations}
Transversality is preserved, and the action is unitary. For $\hat{\bf K}_j(0)$ the analogous results are:
\begin{subequations}
\begin{eqnarray}
&&[\hat{K}_j(0), \hat{a}_l({\bf k})^\dagger]
=-i\dfrac{\hbar}{c}{\sqrt{\omega}}\left(\delta_{lm}-\dfrac{k_lk_m}{|{\bf k}|^2}\right)
 \tilde{\partial}_j(\sqrt{\omega}\ \hat{a}_m({\bf k})^\dagger)\nonumber\\
 &&= -i\dfrac{\hbar}{c}\sqrt{\omega} \tilde{\partial}_j(\sqrt{\omega}\hat{a}_l({\bf k})^\dagger)-\dfrac{i\hbar c}{\omega}{k_l}\hat{a}_j({\bf k})^\dagger;\label{boost1}\\
 &&|{\bf v}|<<c:~~~~~\left(1-\dfrac{i}{\hbar c}{\bf v}\cdot\hat{{\bf K}}(0)\right)|{\bf v}\rangle\simeq
 |{\bf v}+\delta{\bf v}\rangle,\nonumber\\
 &&\delta{ v}_l({\bf k})=\dfrac{1}{c^2}\left(\delta_{lm}-\dfrac{k_lk_m}{|{\bf k}|^2}\right)\sqrt{\omega}\ {\bf v}\cdot 
 \tilde{\mbox{\boldmath$\nabla$}}(\sqrt{\omega}\ { v}_m({\bf k}));\nonumber\\\label{boost2}\\
 && \delta({\bf v}({\bf k})^\ast\cdot{\bf v}({\bf k}))= \tilde{\mbox{\boldmath$\nabla$}}\cdot({\bf v}\dfrac{\omega}{c^2}
 {\bf v}({\bf k})^\ast\cdot{\bf v}({\bf k})).\label{boost3}
  \end{eqnarray}\label{boost}
 \end{subequations}

 This expression for $\delta{\bf v}({\bf k})$ agrees exactly with what one would get from eq.~(\ref{deltaA_deltaE}) for the 
 change $\delta{\bf A}^{(+)}({\bf x})$ in the vector potential caused by an infinitesimal Lorentz transformation, if re-expressed using 
 eq.~(\ref{AE_fromvk}) as a change in ${\bf v}({\bf k})$.

 For later use we carry the discussion of $\hat{\bf J}$ a little further. Equation (\ref{rotation2}) can be expressed as
 \begin{equation}
 \mbox{\boldmath$\alpha$}\cdot\hat{{\bf J}}\ {\bf v}({\bf k})=-i\hbar\ \mbox{\boldmath$\alpha$}\cdot{\bf k}\wedge\tilde{\mbox{\boldmath$\nabla$}}\ {\bf v}({\bf k})+i\hbar\ \mbox{\boldmath$\alpha$}\wedge{\bf v}({\bf k}),
 \label{AM1}
\end{equation}
and in particular with ${\bf k}=k(\sin \theta\cos \varphi, \sin \theta \sin \varphi, \cos \theta)$,
\begin{equation}
\hat{{J}}_3\ v_j({\bf k})=-i\hbar\dfrac{\partial}{\partial\varphi}v_j({\bf k})-i\hbar\ \epsilon_{3jl}\ v_l({\bf k}).
\label{AM2}
\end{equation}
Since $\hat{\bf J}$ obeys the CR's in eq.~(\ref{SO_31}), this infinitesimal action integrates to an action for finite rotations. The result is:
 \begin{eqnarray}
 &&{\rm e}^{-\dfrac{i}{\hbar}{\mbox{\boldmath$\alpha$}}\cdot \hat{\bf J}}|{\bf v}\rangle =| {\bf v}^\prime\rangle,\nonumber\\
 && {\bf v}^\prime({\bf k})=R(\mbox{\boldmath$\alpha$})^{-1}\  {\bf v}(R(\mbox{\boldmath$\alpha$}){\bf k}),\nonumber\\
&& R_{jl}(\mbox{\boldmath$\alpha$})=\delta_{jl}\cos\alpha +\alpha_j\alpha_l \dfrac{(1-\cos\alpha)}{\alpha^2}-\epsilon_{jlm}\ \alpha_m\dfrac{\sin\alpha}
{\alpha},\nonumber\\
&&\alpha= |\mbox{\boldmath$\alpha$}|.
\label{AM3}
\end{eqnarray}
The simultaneous eigenfunctions of $\hat{\bf J}^2$ and
$\hat{J}_3$ (within ${\cal H}_1$) are well
known~\cite{blatt-book,devaney-jmp-74}.
For each pair of integers $(l, m)$ with $l=1, 2, 3, \ldots,
-l\le m\le l$, we have two orthonormal eigenfunctions
depending on $\hat{\bf k}$ alone:
  \begin{subequations}
  \begin{eqnarray}
 && \tilde{\bf L}=-i{\bf k}\wedge \tilde{\mbox{\boldmath$\nabla$}}:\nonumber\\
 && {\bf Y}^{(1)}_{lm}(\hat{{\bf k}})=\dfrac{1}{\sqrt{l(l+1)}} \tilde{{\bf L}}{Y}_{lm}(\hat{{\bf k}}),\nonumber\\
  &&  {\bf Y}^{(2)}_{lm}(\hat{{\bf k}})=\dfrac{1}{\sqrt{l(l+1)}}\hat{{\bf k}}\wedge \tilde{{\bf L}}{Y}_{lm}(\hat{{\bf k}});\label{harmonic1}\\
 && \{\hat{{\bf J}}^2,~\hat{{\bf J}}_3\} {\bf Y}^{(a)}_{lm}(\hat{{\bf k}})=\{\hbar^2l(l+1),m\hbar\} {\bf Y}^{(a)}_{lm}(\hat{{\bf k}}), \nonumber\\&&\qquad a=1, 2;\label{harmonic2}\\
  &&\int_{\mathbb{S}^2} d\Omega(\hat{{\bf k}}) {\bf Y}^{(a^\prime)}_{l^\prime m^\prime}(\hat{{\bf k}})^\ast\cdot {\bf Y}^{(a)}_{lm}(\hat{{\bf k}})
  =\delta_{a^\prime,a}\delta_{l^\prime,l}\delta_{m^\prime,m}.\nonumber\\\label{harmonic3}
  \end{eqnarray}\label{harmonic}
  \end{subequations}
    (Here ${Y}_{lm}(\hat{{\bf k}})$ are the usual spherical harmonics). The two sets of eigenfunctions differ in their parity properties:
  \begin{equation}
  {\bf Y}^{(1)}_{lm}(-\hat{{\bf k}})=(-1)^l{\bf Y}^{(1)}_{lm}(\hat{{\bf k}}), \quad
  {\bf Y}^{(2)}_{lm}(-\hat{{\bf k}})=(-1)^{l+1}{\bf Y}^{(2)}_{lm}(\hat{{\bf k}}).
  \label{parity}
  \end{equation}
  A general ${\bf v}({\bf k})$ has an expansion with two sets of `radial' functions:
   \begin{equation}
   {\bf v}({\bf k})=\sum_{a=1}^{2}\sum_{l=1}^{\infty}\sum_{m=-l}^{l}f_{a,lm}(k){\bf Y}^{(a)}_{lm}(\hat{{\bf k}}),
   \label{expansion}
  \end{equation}
  and the squared norm (\ref{setM}) is
\begin{equation}
 ({\bf v},{\bf v})=\int d^3k~{\bf v}({\bf k})^\ast\cdot{\bf v}({\bf k})=\int_{0}^{\infty} k^2dk \ \sum_{a, l, m}|f_{a,lm}(k)|^2
 \label{norm}
\end{equation}

Finally we treat the operators $\hat{\bf L}, \hat{\bf S}$ in
a similar manner. Following the pattern of eqs.~(\ref{rotation}), for $\hat{L}_j$ we have:
\begin{subequations}
\begin{eqnarray}
&&[\hat{L}_j,\hat{a}_l({\bf k})^\dagger]=i\hbar\{({\bf k}\wedge \tilde{\mbox{\boldmath$\nabla$}})_j\ \hat{a}_l({\bf k})^\dagger
+\dfrac{k_l}{|{\bf k}|^2}( {\bf k}\wedge \hat{{\bf a}}({\bf k})^\dagger)_j\};\nonumber\\\label{OAM1}\\
 &&|\mbox{\boldmath$\alpha$}|<<1:~~~~~\left(1-\dfrac{i}{\hbar}\mbox{\boldmath$\alpha$}\cdot\hat{{\bf L}}\right)|{\bf v}\rangle\simeq
 |{\bf v}+\delta{\bf v}\rangle, \nonumber\\
  &&\delta{\bf v}({\bf k})=-\mbox{\boldmath$\alpha$}\cdot({\bf k}\wedge\tilde{\mbox{\boldmath$\nabla$}}){\bf v}({\bf k})
  -\dfrac{\bf k}{|{\bf k}|^2}\mbox{\boldmath$\alpha$}\cdot{\bf k}\wedge {\bf v}({\bf k});\label{OAM2}\\
 &&\delta({\bf v}({\bf k})^\ast\cdot {\bf v}({\bf k}))=\tilde{\mbox{\boldmath$\nabla$}}\cdot({\bf k}\wedge \mbox{\boldmath$\alpha$}\ {\bf v}({\bf k})^\ast\cdot{\bf v}({\bf k})).~\label{OAM3}
 \end{eqnarray}\label{OAMS}
 \end{subequations}
 These too are consistent with transversality and unitarity.

The case of $\hat{\bf S}$ is particularly interesting, where we find:
\begin{subequations}
\begin{eqnarray}
&&[\hat{S}_j,\hat{a}_l({\bf k})^\dagger]=i\hbar\{ \epsilon_{jlm}\hat{a}_m({\bf k})^\dagger
-\dfrac{k_l}{|{\bf k}|^2}( {\bf k}\wedge \hat{{\bf a}}({\bf k})^\dagger)_j\};\nonumber\\\label{spin1}\\
 &&|\mbox{\boldmath$\alpha$}|<<1:~~~~~\left(1-\dfrac{i}{\hbar}\mbox{\boldmath$\alpha$}\cdot\hat{{\bf S}}\right)|{\bf v}\rangle\simeq
 |{\bf v}+\delta{\bf v}\rangle,\nonumber\\
  &&\delta{\bf v}({\bf k})=\mbox{\boldmath$\alpha$}\wedge{\bf v}({\bf k})
  -\dfrac{\bf k}{|{\bf k}|^2}{\bf k}\cdot\mbox{\boldmath$\alpha$}\wedge {\bf v}({\bf k}) =\mbox{\boldmath$\alpha$}\cdot \hat{\bf k}\ \hat{\bf k}\wedge {\bf v}({\bf k});\nonumber\\\label{spin2}\\
 &&\delta({\bf v}({\bf k})^\ast\cdot {\bf v}({\bf k}))=0.\label{spin3}
 \end{eqnarray}\label{spin}
 \end{subequations}
 (In passing we note that eqs.~(\ref{OAM1}, \ref{spin1})
together give eq.~(\ref{rotation1}), and eqs (\ref{OAM2},
\ref{spin2}) give eq.~(\ref{rotation2}).) In contrast to the
actions of $\hat{\bf J}, \hat{\bf K}(0)$ and $\hat{\bf L}$,
in eqs.~(\ref{spin}) there are no derivatives 
 with respect to ${\bf k}$, the expression for $\delta{\bf v}({\bf k})$ being purely algebraic in ${\bf v}({\bf k})$. This makes the action 
 of $\hat{\bf S}$ on ${\bf v}({\bf k})$ much simpler. Using eq.~(\ref{AM3}) for general $R(\mbox{\boldmath$\alpha$})\in SO(3)$, we have:
 \begin{equation}
|\mbox{\boldmath$\alpha$}|\ll 1: \quad {\bf v}({\bf k})+\delta {\bf v}({\bf k})\simeq R(\mbox{\boldmath$\alpha$}\cdot \hat{\bf k}~\hat{\bf k}){\bf v}({\bf k}).
\label{photospin1}
\end{equation}
Therefore at each $\hat{\bf k}\in \mathbb{S}^2, {\bf v}({\bf k})$ experiences an infinitesimal right handed rotation by a variable $\hat{\bf k}$-dependent angle, 
$\mbox{\boldmath$\alpha$}\cdot \hat{\bf k}$, about the axis $\hat{\bf k}$. For general $\mbox{\boldmath$\alpha$}\in \mathbb{R}^3$, we have:
\begin{equation}
      \mbox{\boldmath$\alpha$}\cdot \hat{{\bf S}}\ {\bf v}({\bf k})=i\hbar \mbox{\boldmath$\alpha$}\cdot \hat{\bf k}\ \hat{\bf k}\wedge {\bf v}({\bf k}).
      \label{photospin2}
\end{equation}
Taking a unit vector $\hat{\mbox{\boldmath$\alpha$}}$ for ${\mbox{\boldmath$\alpha$}}$, we then get
\begin{equation}
(\hat{\mbox{\boldmath$\alpha$}}\cdot \hat{{\bf S}})^2\ {\bf v}({\bf k})=\hbar^2 (\hat{\mbox{{\boldmath$\alpha$}}}\cdot \hat{\bf k})^2 {\bf v}({\bf k}).
\label{photospin3}
\end{equation}
We can exhibit a link between $\hat{{\bf S}}$ and $\hat{{\bf
J}}$ as well. From eqs.~(\ref{AM1}) we find
\begin{equation}
      \hat{\bf k}\cdot\hat{{\bf J}}\ {\bf v}({\bf k})=i\hbar\ \hat{\bf k}\wedge {\bf v}({\bf k})
      \label{hel1}
\end{equation}
implying (on ${\cal H}_1$!)
\begin{equation}
      \hat{{\bf S}} =\hat{\bf k}\ \hat{\bf k}\cdot\hat{{\bf J}}, \quad \hat{{\bf S}}\cdot \hat{{\bf S}}=\hbar^2.
      \label{hel2}
\end{equation}
Since $\hat{\bf k}$ and $\hat{\bf k}\cdot\hat{{\bf J}}$ commute, the commutativity of the components $\hat{S}_j$ is obvious.

It can be seen from these considerations that for each $\hat{\mbox{\boldmath$\alpha$}}, \hat{\mbox{\boldmath$\alpha$}}\cdot \hat{{\bf S}}$ 
has continuous eigenvalues in the interval $(-\hbar, \hbar)$, so it has no normalisable eigenvectors. To simplify the action of 
$\hat{\mbox{\boldmath$\alpha$}}\cdot \hat{\bf S}$ on ${\bf v}({\bf k})$ at each $\hat{\bf k}$, we need to introduce as a particular case of 
eqs.~(\ref{transverse}) an orthonormal pair of transverse `circular polarization' vectors $\mbox{\boldmath$\epsilon$}^{(a)}({\bf k}),\ a=\pm $, obeying:
\begin{eqnarray}
&&\mbox{\boldmath$\epsilon$}^{(a)}({\bf k})^\ast\cdot
\mbox{\boldmath$\epsilon$}^{(b)}({\bf k}) =\delta_{a,
b};\nonumber\\
&&\hat{\bf k}\cdot \mbox{\boldmath$\epsilon$}^{(a)}({\bf
k})=0,\quad \mbox{\boldmath$\epsilon$}^{(-)}({\bf
k})=i\mbox{\boldmath$\epsilon$}^{(+)}({\bf k})^\ast;
\nonumber\\
&&\hat{\bf k}\wedge \mbox{\boldmath$\epsilon$}^{(\pm)}({\bf
k})=\mp i \mbox{\boldmath$\epsilon$}^{(\pm)}({\bf k}),\quad
\mbox{\boldmath$\epsilon$}^{(+)}({\bf k})\wedge
\mbox{\boldmath$\epsilon$}^{(-)}({\bf k})=\hat{\bf
k}.\nonumber\\
\label{CP1}
\end{eqnarray}
If this were possible then we would have
\begin{equation}
\hat{\mbox{\boldmath$\alpha$}}\cdot \hat{\bf S}\
\mbox{\boldmath$\epsilon$}^{(\pm)}({\bf k})=\pm\hbar\
\hat{\mbox{\boldmath$\alpha$}}\cdot \hat{\bf k}\
\mbox{\boldmath$\epsilon$}^{(\pm)}({\bf k}).
\label{CP2}
\end{equation}
(These $\mbox{\boldmath$\epsilon$}^{(a)}({\bf k})$ would
not, however, be eigenvectors of 
$\hat{\mbox{\boldmath$\alpha$}}\cdot \hat{\bf S}$ in ${\cal H}_1$!).

However such $\mbox{\boldmath$\epsilon$}^{(\pm)}({\bf k})$
cannot be found in a globally smooth manner for all
$\hat{\bf k}\in \mathbb{S}^2$.  (If such a choice existed,
then by eqs.~(\ref{CP1}) the real part of
$\mbox{\boldmath$\epsilon$}^{(+)}({\bf k})$ would be a
nowhere vanishing globally smooth tangent vector to
$\mathbb{S}^2$ at $\hat{\bf k}$ for all $\hat{\bf k}\in
\mathbb{S}^2$: but this is not allowed by the `hairy ball'
theorem whose proof is well known~\cite{eisenberg-amm-79}.)
Fortunately, for applications to the paraxial situation in
Section~V, and for the physical interpretation of $\hat{\bf
S}$ developed below, this does not matter. A choice
well-defined everywhere except along the negative $z$-axis
is:

\begin{eqnarray}
&&0\le\theta <\pi, 0\le \varphi<2\pi: \nonumber\\
&& \{\mbox{\boldmath$\epsilon$}^{(+)}({\bf k}),
\mbox{\boldmath$\epsilon$}^{(-)}({\bf
k})\}\nonumber\\&&=R_3(\varphi)R_2(\theta)
R_3(\varphi)^{-1}\left\{\dfrac{1}{\sqrt{2}}\left(\begin{array}{c}1\\i\\0\end{array}\right),
\dfrac{1}{\sqrt{2}}\left(\begin{array}{c}i\\1\\0\end{array}\right)\right\},
\nonumber\\
&&R_3(\varphi)=\left(\begin{array}{ccc}\cos \varphi&
-\sin\varphi&0\\ \sin\varphi&\cos\varphi&
0\\0&0&1\end{array}\right),\nonumber\\
&&R_2(\theta)=\left(\begin{array}{ccc}\cos \theta&
0&\sin\theta\\0&1&0\\-\sin\theta&0&\cos\theta\end{array}\right).
\label{RCPLCP}
\end{eqnarray}
Expanding ${\bf v}({\bf k})$ in this basis as
\begin{equation}
{{\bf v}}({\bf k})=\sum_{a=\pm} \mbox{\boldmath$\epsilon$}^{(a)}({\bf k})v_a({\bf k})
\label{exp1}
\end{equation}
and using eqs.~(\ref{photospin1}, \ref{CP1}) we see that
\begin{equation}
\hat{S}_j\ v_a({\bf k})=a\hbar\ \hat{k}_jv_a({\bf k}), \quad a=\pm.
\label{helicities}
\end{equation}
Therefore in the normalized state $\sqrt{\hbar}|{\bf v}\rangle/\|{\bf v}\|$ the expectation value of $\hat{S}_j$ is
\begin{eqnarray}
\dfrac{\hbar}{\|{\bf v}\|^2} \langle {\bf v}|\hat{S}_j|{\bf
v}\rangle&=&\hbar \int d^3 k\ \hat{k}_j (p({\bf k}, +)- p({\bf k}, -)),\nonumber\\
p({\bf k}, a)&=&|v_a({\bf k})|^2/\|{\bf v}\|^2,
 \label{spin_average}
\end{eqnarray}
where $p({\bf k}, \pm)$ are the probability densities in
${\bf k}$-space for the photon to have momentum $\hbar {\bf
k}$ and to be 
right/left circularly polarized.

This discussion helps to bring out the meaning of the
operators $\hat{S}_j$ as single photon observables. In
particular the operators $\hat{S_j}$ (within single photon
subspace) are completely defined by the expectation value
expression~(\ref{spin_average}) since it is given for all possible
states. As is evident these operators commute with the
momentum operator for the photon.
\section{The paraxial approximation, relation to
quantization procedure}
The treatment of the paraxial regime is an important part of
classical ray as well as wave optics, which is important
also for laser physics. 
In this Section we examine the problem of combining it in a
physically reasonable way with the quantization of the
radiation field outlined in the 
previous Section.
\subsection*{Scalar paraxial case}
For simplicity we outline first the sequence of assumptions
and approximations involved in paraxial optics (in leading
order)
in the scalar context, and then turn to the transverse
vector potential. We are concerned with complex scalar
analytic signal solutions 
 $\psi(x)$ to the wave equation
\begin{equation}
\left( \mbox{\boldmath$\nabla$}^2-\dfrac{1}{c^2}\dfrac{\partial^2}{\partial t^2}\right) \psi(x)=0.\label{Wave_Eqn}
\end{equation}
Expressing a general solution as a Fourier integral over positive frequencies,
\begin{equation}
 \psi(x)=\int_0^\infty d\omega\ \tilde{\psi}({\bf x};\omega)e^{-i\omega t},\label{TimeFourier}
\end{equation}
the function $\tilde{\psi}$ obeys the free Helmholtz equation ($\omega=ck=2\pi c/\lambda$)
\begin{equation}
 \left( \mbox{\boldmath$\nabla$}^2+k^2\right) \tilde{\psi}({\bf x};\omega)=0.\label{Helmhotz}
\end{equation}
The general solution of this equation, as a superposition of plane waves, is
\begin{equation}
\tilde{\psi}({\bf x};\omega)=\int_{\mathbb{S}^2} d\Omega(\hat{{\bf k}})\ \phi({\bf k})\ e^{i{\bf k}\cdot{\bf x}},\quad |{\bf k}|=k,\label{Ang_Spectrum}
\end{equation}
where we may assume $\phi({\bf k})\in L^2(\mathbb{S}^2)$. Now we assume there are only `forward propagating' waves in the sense that $\phi({\bf k})$
vanishes for $k_3<0$. Then
\begin{eqnarray}
&& \tilde{\psi}({\bf x};\omega)=\int_{|{\bf k_\perp}|\leq k} d^2{\bf k_\perp} \dfrac{\phi({\bf k_\perp},\sqrt{k^2-{\bf k_\perp}^2})}{k\sqrt{k^2-{\bf k_\perp}^2}}\nonumber\\&&\times
 e^{i{\bf k_\perp}\cdot {\bf x_\perp}+i\sqrt{k^2-{\bf k_\perp}^2}z}.\label{2dFourier}
\end{eqnarray}

At this point we introduce the paraxial condition -- $\phi({\bf k})$ is negligible unless $|{\bf k_\perp}|\ll k$:
\begin{equation}
 {\phi({\bf k_\perp},\sqrt{k^2-{\bf k_\perp}^2}})\approx 0 ~~\text{unless}~~|{\bf k_\perp}|<<k.\label{Paraxial_condition}
\end{equation}
With this assumption we can carry out two simplifications:
expand the square root in the exponent in eq.~(\ref{2dFourier}) to lowest nontrivial order, and formally
extend the integration region to the entire plane:
\begin{eqnarray}
\tilde{\psi}_{\text{par}}({\bf x};\omega)&\simeq&
e^{ikz}\psi_{0,\text{par}}({\bf
x_\perp},z;\omega),\nonumber\\
\psi_{0,\text{par}}({\bf
x_\perp},z;\omega)&=&\int_{\mathbb{R}^2}d^2{\bf k_\perp}
\dfrac{\phi({\bf k_\perp},k)}{k^2} e^{i{\bf k_\perp}\cdot
{\bf x_\perp}-i\frac{\lambdabar}{2}
 {\bf k_\perp}^2z}.\nonumber\\\label{Paraxial_field}
\end{eqnarray}
The amplitude $\psi_{0,\text{par}}$ obeys the paraxial wave equation (PWE) for  (reduced) wave length $\lambdabar=\lambda/2\pi$:
\begin{equation}
 i\dfrac{\partial}{\partial z}\psi_{0,\text{par}}({\bf x_\perp},z;\omega)=-\dfrac{\lambdabar}{2}\nabla^2_\perp ~\psi_{0,\text{par}}({\bf x_\perp},z;\omega).\label{PWE}
\end{equation}

If we now try to retrace these steps to arrive at an approximate paraxial solution to the wave equation (\ref{Wave_Eqn}), we realize that for the 
condition (\ref{Paraxial_condition}) to make sense the range of frequencies involved in eq.~(\ref{TimeFourier}) must be greater than some 
minimum $\omega_{\text{min}}>0$. Incorporating this, we arrive at the following form for an approximate paraxial scalar analytic signal 
obeying eq.~(\ref{Wave_Eqn}):
\begin{equation}
 \psi_{\text{par}}(x) \simeq c\int_{k_{\text{min}}}^\infty dk\ e^{ik(z-ct)}\ \psi_{0,\text{par}}({\bf x_\perp},z;\omega),\label{Paraxial_signal}
\end{equation}
with $\psi_{0,\text{par}}({\bf x_\perp},z;\omega)$ given by eq.~(\ref{Paraxial_field}) and $\phi({\bf k})$ obeying (\ref{Paraxial_condition}).

At this point it is important to realize that while $\psi_{0,\text{par}}({\bf x_\perp},z;\omega)$ is a paraxial solution of the PWE (\ref{PWE}), 
there are nonparaxial solutions as well. For the moment suppress $\omega$ as an argument. If at, say, $z=0$ we choose as `initial' amplitude a 
general $\psi({\bf x_\perp}, 0)\in L^2(\mathbb{R}^2_{{\bf x_\perp}})$, and express it as
\begin{equation}
{\psi}({{\bf x_\perp}},0)=\int_{\mathbb{R}^2} d^2{\bf k_\perp} ~\phi({\bf k_\perp})\
~e^{i{\bf k_\perp}\cdot {\bf x_\perp}},\label{2dFT}
\end{equation}
then $\phi({\bf k_\perp})$ is a general element of
$L^2(\mathbb{R}^2_{{\bf k_\perp}})$ and the solution of
(\ref{PWE}) is
\begin{equation}
\psi({{\bf x_\perp}},z)=\int_{\mathbb{R}^2} d^2{\bf
k_\perp}~\phi({\bf k_\perp})
~ e^{i{\bf k_\perp}\cdot {\bf
x_\perp}-i\frac{\lambdabar}{2}{\bf k_\perp}
^2z}.\label{PWE_solution}
\end{equation}
This $\phi({\bf k_\perp})$ need not obey
(\ref{Paraxial_condition}), so $\psi({\bf x_\perp}, z)$ may
not be paraxial at all. 
Thus most solutions to the PWE are nonparaxial.

To emphasize this fact, we take this discussion one step
further. Choose {\bf any} discrete orthonormal basis 
$\{\phi_n({\bf k_\perp})\}$ for $L^2(\mathbb{R}^2_{{\bf k_\perp}})$:
\begin{eqnarray}
& \int_{\mathbb{R}^2} d^2{\bf k_\perp}~\phi_{n^\prime}({\bf
k_\perp})^\ast ~\phi_n({\bf
k_\perp})=\delta_{n^\prime,n},~~~n^\prime,n=1,2,\cdots;\nonumber\\
& \sum\limits_{n=1}^\infty~\phi_n({\bf k_\perp}) ~\phi_n({\bf
k_\perp^\prime})^\ast=\delta^{(2)}({\bf k_\perp}-{\bf
k_\perp^\prime}).\label{k_ONB}
\end{eqnarray}
By Fourier transformation we get a corresponding orthonormal
basis for $L^2(\mathbb{R}^2_{{\bf x_\perp}})$, and a set of
PWE solutions:
\begin{subequations}
\begin{eqnarray}
& {\psi}_n({{\bf x_\perp}},0)=\dfrac{1}{2\pi}\int  d^2{\bf
k_\perp} ~\phi_n({\bf k_\perp})
e^{i{\bf k_\perp}\cdot {\bf x_\perp}},\nonumber\\
& \int_{\mathbb{R}^2} d^2{\bf x_\perp}~\psi_{n^\prime}({\bf
x_\perp},0)^\ast ~\psi_n({\bf
x_\perp},0)=\delta_{n^\prime,n},\nonumber\\
& \sum_{n=1}^\infty~\psi_n({\bf x_\perp},0) ~\psi_n({\bf
x_\perp^\prime}, 0)^\ast=\delta^{(2)}({\bf x_\perp}-{\bf
x_\perp^\prime});\nonumber\\\label{x_ONB1}\\
& \psi_n({{\bf
x_\perp}},z)=\dfrac{1}{2\pi}\int_{\mathbb{R}^2} d^2{\bf
k_\perp}~\phi_n({\bf k_\perp})
e^{i{\bf k_\perp}\cdot {\bf
x_\perp}-i\frac{\lambdabar}{2}{\bf k_\perp}
^2z}.\nonumber\\\label{x_ONB2}
\end{eqnarray}\label{x_ONB}
\end{subequations}
For each $z, \{\psi_n({{\bf x_\perp}},z)\}$ is an
orthonormal basis for $L^2(\mathbb{R}^2_{{\bf x_\perp}})$.
Combining eqs.~(\ref{k_ONB},\ref{x_ONB1}) 
also gives
\begin{equation}
e^{i{\bf k_\perp}\cdot {\bf
x_\perp}-i\frac{\lambdabar}{2}{\bf k_\perp}
^2z}=2\pi\sum_{n=1}^{\infty}\phi_n({\bf
k_\perp})^\ast\psi_n({\bf x_\perp},z).\label{PWE_identity}
\end{equation}
Now a general initial $\psi({\bf x_\perp}, 0)$, eq.~(\ref{2dFT}), can be expanded as
\begin{equation}
\psi({\bf x_\perp},0)=\sum_{n=1}^\infty c_n \psi_n({\bf
x_\perp},0),\label{ONB_expansion}
\end{equation}
where $\{c_n\}$ is an $l^2$-sequence,
\begin{equation}
\sum\limits_{n=1}^\infty |c_n|^2 <  \infty,\label{l_condition}
\end{equation}
{\bf which is otherwise unrestricted}. This initial
$\psi({\bf x_\perp},0)$ then evolves to the solution
\begin{equation}
\psi({\bf x_\perp},z)=\sum_{n=1}^\infty c_n \psi_n({\bf
x_\perp},z)\label{PWE_ONB}
\end{equation}
to the PWE {\bf which may not be paraxial at all}. If
$\psi({\bf x_\perp},z)$ is to be a {\bf paraxial} solution
to the PWE, then 
$\sum_{n=1}^\infty c_n \phi_n({\bf k_\perp})$ must obey
(\ref{Paraxial_condition}), which implies conditions on
$\{c_n\}$ going well beyond 
the $l^2$ property~(\ref{ONB_expansion}).
\subsection*{Paraxial vector potential, route to quantization}
The general form of the complex analytic signal transverse
vector potential is given in eq.~(\ref{AE_fromvk}) in terms
of ${\bf v}({\bf k})$. 
For ${\bf A}^{(+)}(x)$ to be paraxial, we see now that in
addition to ${\bf k}\cdot {\bf v}({\bf k})=0$, we must have
\begin{eqnarray}
{\bf v}({\bf k_\perp},k_3)=0~\text{for}\ \ k\leq
k_{\text{min}}, ~k_3~<~0;\nonumber\\
{\bf v}({\bf k_\perp},\sqrt{k^2-{\bf k_\perp}^2})\approx
0~\text{unless}\ \ |{\bf
k_\perp}|<<k.\label{vector_paraxial}
\end{eqnarray}
Given such ${\bf v}({\bf k})$, and following steps as in the
scalar case, we get the following approximate form for a
paraxial vector potential:
\begin{eqnarray}
&&{\bf A}^{(+)}_{\text{par}}(x)\simeq
\dfrac{c}{2\pi}\int_{k_{\text{min}}}^{\infty}
~k~dk~e^{ik(z-ct)}\int_{\mathbb{R}^2}~\dfrac
{d^2k_{\perp}}{\sqrt{\omega}\sqrt{k^2-{\bf
k_\perp}^2}}\nonumber\\&&\times {\bf v}({\bf k_\perp},k)
e^{i{\bf k_\perp}\cdot {\bf
x_\perp}-i\frac{\lambdabar}{2}{\bf k_\perp}
^2z}.\label{paraxial_potential}
\end{eqnarray}
Turning to `quantization in the paraxial regime', to begin
with it may seem reasonable to proceed as follows: start
with the form~(\ref{paraxial_potential}) for the classical vector
potential in this regime, then proceed to `quantize' it by
turning  ${\bf A}^{(+)}_{\rm par}(x)$ into 
a suitable operator $\hat{\bf A}^{(+)}_{\rm par}(x)$. This
would mean taking the paraxial approximation first, then
performing quantization. For instance, 
in the spirit of eqs.~(\ref{k_ONB}~\ref{PWE_ONB}), we may
expand ${\bf A}^{(+)}_{\rm par}(x)$ in some complete
orthonormal set 
$\{\mbox{\boldmath$\psi$}_n({\bf x_\perp}, 0)\}$, each
member of which is paraxial, and obtain expansion
coefficients similar to 
$\{c_n\}$ in eq.~(\ref{ONB_expansion}); and then convert
them to operators as $c_n\rightarrow\hat{a_n},
c_n^\ast\rightarrow \hat{a_n}^\dagger$. 
However this procedure  seems inadvisable for important
physical reasons:
\begin{itemize}
\item[(i)] The paraxial regime is defined only in an approximate
way, as seen in eqs.(\ref{Paraxial_condition}, \ref{Paraxial_signal},
\ref{vector_paraxial}, \ref{paraxial_potential}). It is not
possible to define it with any degree of 
mathematical precision, such as is associated with the quantization process.
\item[(ii)] In effect, only the classical amplitudes ${\bf v}({\bf
k_\perp}, k_3)$ for $k\ge k_{\rm min}> 0, k_3>0, |{\bf
k_\perp}|\ll k$, would be converted 
into operators.
\item[(iii)] Even if this were possible, the classical condition
(\ref{vector_paraxial}) would have to be translated into
statements on the resulting 
operators in some way, not however on the CCR's themselves.
This is in essence the problem of restrictions on $\{c_n\}$
mentioned after 
eq.~(\ref{PWE_ONB}).
\end{itemize}
For these reasons, a better procedure seems to be as
follows. We begin with the quantized radiation field as set
up in Section~III, in which with every 
classical analytic signal vector potential ${\bf
A}^{(+)}(x)$ we are able to associate an (unnormalized)
photon annihilation-creation operator 
pair $\hat{a}({\bf v}), \hat{a}({\bf v})^\dagger$ obeying
eqs.~(\ref{single_photon}). Here ${\bf v}({\bf k})$ is any
element of the classical 
Hilbert space ${\cal M}$, eq.~(\ref{setM}). The operator
$\hat{a}({\bf v})^\dagger$ creates photons in the state, or
with wave function, 
${\bf v}({\bf k})$. We then limit the choice of ${\bf
v}({\bf k})$ to those obeying the paraxial conditions
(\ref{vector_paraxial}), so we deal 
with a limited subset of the operator pairs $\hat{a}({\bf
v}), \hat{a}({\bf v})^\dagger$. In this way, the paraxial
conditions come after the 
quantization of the entire field, rather than the other way
around. We accept the approximate nature of the statement of
the paraxial conditions; 
and that ${\bf A}^{(+)}_{\rm par}(x)$ of
(\ref{paraxial_potential}) obeys the wave equation only
approximately. This understandable lack of 
precision gets expressed in the properties of the chosen
${\bf v}({\bf k})$'s, not in the quantization process.
However, The procedure adopted here differs from that
in~\cite{deutsch-pra-91}, wherein an attempt is made to
identify a subspace
of ${\mathcal H}_1$, made up
of paraxial wavefunctions. It would have to be checked if
such a space is a closed as well as a proper subspace of
${\mathcal H} _1$.
In reference~\cite{calvo-pra-06}, on the other hand, the paraxial limit
is defined by an inequality ${\mathcal \theta} << 1$
for a parameter ${\mathcal \theta}$  that governs the degree
of paraxiality, and quantization is done after this limit is
taken in the classical vector potential. There is therefore
a vagueness as to which original classical amplitudes are 
being made into operators.

In our approach, there is no 
paraxial vector potential field operator $\hat{\bf
A}^{(+)}_{\rm par}(x)$ at all. An examination of some
properties of one photon states 
$|{\bf v}\rangle$ for paraxial ${\bf v}({\bf k})$ is taken
up in the next Section.

We conclude this Section by mentioning the
Laguerre--Gaussian mode functions which are a particular
widely used example of the complete orthonormal 
sets $\{\phi_n({\bf k_\perp})\}, \{\psi_n({\bf x_\perp},
z)\}$ in eqs.~(\ref{k_ONB}--\ref{PWE_identity}).
They are characterized by a real waist 
parameter $w>0$; and the index $n=1, 2, \ldots$ is replaced
by a pair $(m, p), m=0, \pm 1, \pm 2, \ldots$ and $p=0, 1,
2, \ldots$ independently. A readable account is available in
reference~\cite{Pampaloni-arxiv-04}.

(The index $m$ used here -- the magnetic quantum number --
is in the usual notation of the quantum theory of angular
momentum (QTAM). It is often 
replaced by $l$ in the literature, which however has a
different meaning in QTAM). Then the 
$\phi_n$'s are, with ${\bf k_\perp}=\rho (\cos \varphi, \sin \varphi)$:
\begin{eqnarray}
 &\phi_{m,p}({\bf k_\perp})=\dfrac{w}{\sqrt{2\pi}}\sqrt{\dfrac{p!}{(p+|m|)!}}~
 ~e^{im\varphi}\left(\dfrac{iw\rho}{\sqrt{2}}\right)^{|m|}\nonumber\\&\times
 L_p^{|m|}\left(\dfrac{w^2\rho^2}{2}
 \right)e^{-\dfrac{w^2\rho^2}{4}}.\label{k_LG}
\end{eqnarray}
Here the $L$'s are Laguerre polynomials. For the $\psi_n$'s we need the expressions
\begin{eqnarray}
 &w({\zeta})=w(1+\zeta^2)^{1/2},\quad  \zeta= z/z_R,\nonumber\\
&z_R=w^2/2\lambdabar=~\text{Rayleigh range}.\label{Rayeliegh}
\end{eqnarray}
Then with ${\bf x_\perp}=r(\cos \phi, \sin \phi)$, the $\psi_n$'s are:
\begin{eqnarray}
 &&\psi_{m,p}({\bf x_\perp},z)=\dfrac{(-1)^{p+|m|}}{w(\zeta)}\sqrt{\dfrac{2}{\pi}\dfrac{p!}{(p+|m|)!}}\nonumber\\
 &&\times e^{-i(2p+|m|+1)\tan^{-1}\zeta}e^{im\phi}\left(\dfrac{\sqrt{2}r}{w(\zeta)}\right)^{|m|}\nonumber\\
 &&\times L_p^{|m|}~~\left(\dfrac{2r^2}{w(\zeta)^2}
 \right)~~e^{-\dfrac{r^2}{w^2(1+i\zeta)}}.\label{x_LG}
\end{eqnarray}
For any $w>0$, the $\psi_{m,p}({\bf x_\perp},z)$ are solutions of the PWE (\ref{PWE}); to be paraxial solutions we must require $w\gg \lambdabar$. The 
identity (\ref{PWE_identity}) now reads:
\begin{eqnarray}
 e^{i{\bf k_\perp}\cdot {\bf x_\perp}-i\frac{\lambdabar}{2}{\bf k_\perp^2}z}&=&2\pi\sum_{p=0,1,\cdots}\sum_{m=0,\pm 1,\cdots}\phi_{m,p}({\bf k_\perp})^\ast\nonumber\\
 &&\times \psi_{m,p}({\bf x_\perp},z).\label{PWE_LG}
 \end{eqnarray}
The Gaussian factor in $\phi_{m,p}({\bf k_\perp})$ ensures
paraxiality but at different rates for different $m, p$. Any
finite linear combination of the $\phi_{m,p}({\bf k_\perp})$
is also paraxial, so such a combination of $\psi_{m,p}({\bf
x_\perp},z)$ is a paraxial solution of the PWE. However this
is not necessarily true if we choose an otherwise
unrestricted $l^2$-sequence $\{C_{m, p}\}$ and form the
corresponding combination of $\psi_{m,p}({\bf x_\perp},z)$.
This is an instance of the comments made after
eq.~(\ref{PWE_ONB}).

We may repeat that the L-G mode functions are only one
example of a complete orthonormal set along the lines  of
eqs.~(\ref{k_ONB}, \ref{x_ONB}), 
though important for practical purposes. While these
functions have been recalled here in the scalar case, the
extension to the transverse vector 
 case is taken up in the next Section.
\section{Vector potential and single photon states in the
paraxial regime}
As recalled in Section 3, the hermitian operators $\hat{\bf
L}$ and $\hat{\bf S}$, unlike $\hat{\bf J}$, do not obey the
CR's of a quantum mechanical 
angular momentum. Therefore, while from eq.~(\ref{harmonic2}) the eigenvalues of $\hat{\bf J}^2,
\hat{J}_3$ within ${{\cal H}_1}$ are known to have the 
familiar quantized forms $\hbar^2 l(l+1)$, $m\hbar$ with
$l=1, 2, \ldots,$ and $-l\le m\le l$, there is no reason to
expect similar patterns for the 
eigenvalues of $\hat{\bf L}^2, \hat{L}_3$ or $\hat{\bf S}^2,
\hat{S}_3$. Indeed, we have found that the $\hat{S}_j$ can
be simultaneously diagonalised 
and, within ${\cal H}_1$, each $\hat{S}_j$ has continuous
eigenvalues in the interval $(-\hbar, \hbar)$ while obeying
$\hat{S}_j\hat{S}_j=\hbar^2$.

Now, to carry forward the analysis of Section 4 and
understand better the properties of the paraxial regime, we
first examine the eigenfunctions 
of $\hat{J}_3$ within ${\cal H}_1$. We use spherical polar
or cylindrical variables as convenient, and consider both
${\bf k}$-space and physical 
space expressions.

\subsection*{Forms of $\hat{J}_3$ eigenfunctions in ${\cal H}_1$}
The action of $\hat{J}_3$ on a wave function ${\bf v}({\bf
k})\in {\cal M}$ is given in eq.~(\ref{AM2}). Since
$\hat{J}_3$ has discrete eigenvalues, 
normalisable eigenfunctions can be constructed. The general
solution (in spherical polar variables ${\bf k}\rightarrow
k, \theta, \varphi$) to
\begin{eqnarray}
&&\hat{{J}}_3{\bf v}_m(k, \theta, \varphi)=m\hbar{\bf v}_m
(k, \theta, \varphi),\quad m=0, \pm 1, \pm 2,
\ldots,\nonumber\\
&&{\bf k}\cdot {\bf v}_m(k, \theta, \varphi)=0,
\label{J3_problems}
\end{eqnarray}
is a linear combination of
\begin{eqnarray}
&\mbox{\boldmath$\alpha$}_m(\theta, \varphi)=
e^{i(m-1)\varphi}
\left(\begin{array}{c}C\\iC\\-S~e^{i\varphi}\end{array}\right),\nonumber\\
 &
\mbox{\boldmath$\beta$}_m(\theta, \varphi)=
e^{i(m+1)\varphi}
\left(\begin{array}{c}iC\\C\\-iS~e^{-i\varphi}\end{array}\right),
\nonumber\\
&C=\cos \theta, S=\sin \theta,\label{J3_eigenfns}
\end{eqnarray}
with any functions of $k,\theta$ as coefficients (subject to
${\bf v}_m(k, \theta, \varphi)$ being singlevalued at
$\theta=0, \pi$). 
The column vectors here are however not mutually orthogonal.
Using the circular polarization basis vectors 
$\mbox{\boldmath$\epsilon$}^{(\pm)}(\hat{\bf k})$ of eq.~(\ref{RCPLCP}), we can find an alternative construction. The
orthonormal vectors 
$\mbox{\boldmath$\epsilon$}^{(\pm)}(\hat{\bf k})$ are
\begin{eqnarray}
&\mbox{\boldmath$\epsilon$}^{(+)}(\theta,
\varphi)=\dfrac{e^{i\varphi}}{\sqrt{2}}\begin{pmatrix}\cos
\theta \cos\varphi -i\sin\varphi\cr\cos
\theta\sin\varphi+i\cos\varphi \cr
-\sin\theta\end{pmatrix},\nonumber\\
&\mbox{\boldmath$\epsilon$}^{(-)}(\theta,
\varphi)=i\mbox{\boldmath$\epsilon$}^{(+)}(\theta,
\varphi)^\ast.\label{Poln_basis}
\end{eqnarray}
As expected, at $\theta=\pi$ we have $\varphi$-dependent limits:
\begin{eqnarray}
&\mbox{\boldmath$\epsilon$}^{(+)}(\pi,
\varphi)=\dfrac{e^{2i\varphi}}{\sqrt{2}}\begin{pmatrix}-1\cr
i \cr 0\end{pmatrix},\nonumber\\
&\mbox{\boldmath$\epsilon$}^{(-)}(\pi,
\varphi)=-i\dfrac{e^{-2i\varphi}}{\sqrt{2}}\begin{pmatrix}1\cr
i \cr 0\end{pmatrix}.\label{South_pole}
\end{eqnarray}
We then find that any ${\bf v}_m(k, \theta, \varphi)$
obeying eq.~(\ref{J3_problems}) is a $(k, \theta)$ dependent
linear combination of
\begin{equation}
e^{i(m-1)\varphi}\mbox{\boldmath$\epsilon$}^{(+)}(\theta,
\varphi),\quad
e^{i(m+1)\varphi}\mbox{\boldmath$\epsilon$}^{(-)}(\theta,
\varphi),\label{alter_eigenfns}
\end{equation}
subject again to being singlevalued at $\theta=0, \pi$.

The sets (\ref{J3_eigenfns}), (\ref{alter_eigenfns}) of
$\hat{{J}}_3$ eigenfunctions are linearly related:
\begin{eqnarray}
&& \begin{pmatrix}\mbox{\boldmath$\alpha$}_m(\theta,
\varphi)\cr \mbox{\boldmath$\beta$}_m(\theta,
\varphi)\end{pmatrix}=\dfrac{1}{\sqrt{2}}
\left(\begin{array}{ll}(1+C)&-i(1-C)\cr i(1-C) &
(1+C)\end{array}\right) \nonumber\\
&&\hspace{5pc}\times \begin{pmatrix}e^{i(m-1)\varphi}\
\mbox{\boldmath$\epsilon$}^{(+)}(\theta, \varphi) \cr
e^{i(m+1)\varphi}\ \mbox{\boldmath$\epsilon$}^{(-)}(\theta,
\varphi) \end{pmatrix};\nonumber\\
&&\begin{pmatrix}e^{i(m-1)\varphi}\
\mbox{\boldmath$\epsilon$}^{(+)}(\theta, \varphi) \cr
e^{i(m+1)\varphi}\ \mbox{\boldmath$\epsilon$}^{(-)}(\theta,
\varphi), \end{pmatrix}\nonumber\\
&&=\dfrac{1}{2\sqrt{2}}
\left(\begin{array}{ll}1+\sec\theta & i(\sec\theta-1)\cr
-i(\sec\theta -1) & 1+\sec\theta\end{array}\right)
\begin{pmatrix}\mbox{\boldmath$\alpha$}_m(\theta,
\varphi)\cr \mbox{\boldmath$\beta$}_m(\theta,
\varphi)\end{pmatrix}.\nonumber\\
\label{2_bases_reln}
\end{eqnarray}
Therefore, returning to eq.~(\ref{J3_problems}), we can write in general
\begin{eqnarray}
{\bf v}_m(k, \theta, \varphi) &=&a(k,
\theta)e^{i(m-1)\varphi}\
\mbox{\boldmath$\epsilon$}^{(+)}(\theta, \varphi) + b(k,
\theta)\nonumber\\
&&\times e^{i(m+1)\varphi}\
\mbox{\boldmath$\epsilon$}^{(-)}(\theta,
\varphi),\label{general_eigenfn}
\end{eqnarray}
with $a(k, \theta), b(k, \theta)$ suitably behaved at
$\theta=0, \pi$ but otherwise arbitrary. If ${\bf v}_{m'}(k,
\theta, \varphi)$ is 
another $\hat{J}_3$ eigenfunction for eigenvalue $m'\hbar$
involving $a'(k, \theta), b'(k, \theta)$ we find:
\begin{eqnarray}
&&({\bf v}_{m'}, {\bf v}_{m})=\int d^3k\ {\bf v}_{m'} (k,
\theta, \varphi)^\ast\cdot  {\bf v}_{m} (k, \theta,
\varphi)\nonumber\\
&=&2\pi\delta_{m',m}\int_0^\infty k^2\
dk\int_{0}^{\pi}\sin\theta\ d\theta (a'(k, \theta)^\ast a(k,
\theta)\nonumber\\
 &&+b'(k, \theta)^\ast b(k, \theta)).\label{inner_product}
\end{eqnarray}
Thus, assuming ${\bf v}(k, \theta, \phi)$ is normalized,
$\hat{k} \cdot \hat{\bf S}$ has a
definite expectation value and a finite nonzero spread with
respect to it.
It may be emphasized that the two terms in eq.~(\ref{general_eigenfn}) are not related to eigenfunctions of
$\hat{k}\cdot\hat{\bf S}$ 
(in ${\cal H}_1$) in any way, as these are not normalisable.

The general structure of $\hat{{J}}_3$ eigenfunctions in
physical space, i.e., their azimuthal dependences, can be
found using 
eq.~(\ref{AE_fromvk}). A transverse analytic signal ${\bf
A}^{(+)}(x)$ obeying
\begin{equation}
\hat{{J}}_3{\bf A}^{(+)}(x)=m\hbar{\bf
A}^{(+)}(x)\label{vector_potential1}
\end{equation}
is a complex numerical linear combination of three-component
amplitudes of the form (in spherical polar variables ${\bf
x}\rightarrow r, \theta, \phi)$
\begin{equation}
e^{i(m-1)\phi}\begin{pmatrix}a(r, \theta, t)\cr ia(r,
\theta, t)\cr c(r, \theta, t)e^{i\phi}\end{pmatrix},
e^{i(m+1)\phi}\begin{pmatrix}ib(r, \theta, t)\cr b(r,
\theta, t)\cr c'(r, \theta,
t)e^{-i\phi}\end{pmatrix},\label{vector_potential2}
\end{equation}
subject to the wave equation for ${\bf A}^{(+)}(x)$,
transversality, singlevaluedness and the positive frequency
condition. 
In ${\bf x}$-space there is no analogue to the solutions of
the form (\ref{alter_eigenfns}), 
since $\mbox{\boldmath$\epsilon$}^{(\pm)}({\hat{\bf k}})$
are local in ${\bf k}$-space alone.
\subsection*{Photon wave functions and vector potential in
paraxial limit}
We begin with consequences of eqs.~(\ref{CCR},
\ref{single_photon}):
\begin{eqnarray}
& {\bf v}({\bf k})\in{\cal M}: ~~\hat{a}_j({\bf k})|{\bf
v}\rangle = \frac{1}{\sqrt{\hbar}}v_j({\bf
k})|0\rangle,\nonumber\\
&\langle {\bf v}|\hat{a}_j({\bf k})^\dagger =
\frac{1}{\sqrt{\hbar}}v_j({\bf k})^\ast\langle
0|.\label{1_photon}
\end{eqnarray}
 As discussed in Section 4, in the paraxial case it is the choice of ${\bf v}({\bf k})$ that is suitably restricted, without affecting the 
 operators $\hat{\bf a}({\bf k}), \hat{\bf a}({\bf k})^\dagger$.

 The paraxial region in ${\bf k}$-space consists of wave vectors along and very close to the positive $z$-axis:
 \begin{eqnarray}
 & {\bf k}=({\bf k_\perp},\sqrt{k^2-{\bf k}_\perp^2})= k(\sin\theta\cos\varphi,\sin\theta\sin\varphi,\cos\theta)\nonumber\\
  &\simeq k(\theta\cos\varphi,\theta\sin\varphi,1-\theta^2/2),~~ 0\leq\theta <<1,\label{paraxial_region}
 \end{eqnarray}
 where only terms upto quadratic in $\theta$ are retained in all relevant expressions. The condition on ${\bf v}({\bf k})$ stated in 
 eq.~(\ref{vector_paraxial}) is that it be negligible outside of the region (\ref{paraxial_region}):
 \begin{equation}
   {\bf v}({\bf k})\approx 0~~\text{unless}~\theta<<1,\  {\rm i.e.}\ \ |{\bf k_\perp}|<< k.\label{paraxial_condition1}
 \end{equation}
Then transversality gives
 \begin{equation}
  v_3({\bf k})\simeq -\theta(\cos\varphi ~v_1({\bf k})+\sin\varphi ~v_2({\bf k})).\label{long_component}
  \end{equation}
Qualitatively stated, $v_3({\bf k})$ is one order of magnitude smaller than $v_\perp({\bf k})$.

To apply these considerations to the exact $\hat{J}_3$ eigenfunctions ${\bf v}_m(k, \theta, \varphi)$ in eq.~(\ref{general_eigenfn}), we 
impose the paraxial property on $a(k,\theta),~b(k,\theta)$:
  \begin{equation}
  a(k,\theta),~b(k,\theta) \approx 0~~\text{unless}~\theta<<1.\label{paraxial_condition2}
\end{equation}
We now see that if such a paraxial ${\bf v}_m(k, \theta, \varphi)$, with both $a(k,\theta)$ and $b(k,\theta)$ sharply peaked about $\theta=0$, is 
expanded in the total angular momentum eigenfunctions $\{{\bf Y}^{(a)}_{l,m}(\theta, \varphi)\}$ of eq.~(\ref{harmonic}), on account of the 
uncertainty principle many terms with a large spread of $l$ values will be present. On the other hand, since the paraxial region (\ref{paraxial_region}) 
is preserved under rotations about the $z$-axis, individual $\hat{J}_3$ eigenfunctions remain useful in this regime; the label `$m$' remains a 
`good quantum number'. This explains the motivation to study ${\bf v}_m(k, \theta, \varphi)$ in this limit.

The vector potential in the paraxial limit shows some subtleties. We start with the analogues to the scalar equations 
(\ref{TimeFourier}, \ref{Helmhotz}):
\begin{eqnarray}
 && {\bf A}^{(+)}(x)=\int_{0}^{\infty} d\omega~ \tilde{\bf A}^{(+)}({\bf x};\omega)e^{-i\omega t};\nonumber\\
&& (\mbox{\boldmath$\nabla$}^2+k^2)~\tilde{\bf A}^{(+)}({\bf x};\omega)=0,\quad \mbox{\boldmath$\nabla$}\cdot \tilde{\bf A}^{(+)}({\bf x};\omega)=0.\nonumber\\
\label{vector_potential_frequency}
\end{eqnarray}
When ${\bf v}({\bf k})$ obeys eqs.~(\ref{paraxial_condition1}, \ref{long_component}) there is some minimum $\omega_{\rm min}>0$; and consistent with 
eq.~(\ref{paraxial_potential}) we have:
\begin{equation}
\tilde{\bf A}^{(+)}({\bf x}_\perp, z;\omega)\simeq \dfrac{e^{ikz}}{2\pi\sqrt{\omega}}\int d^2{\bf k}_\perp~{\bf v}({\bf k}_\perp, k)e^{i{\bf k}_\perp\cdot {\bf x}_\perp-i\frac{\lambdabar}{2}{\bf k}_\perp^2z}.
\label{paraxial_vector_potential}
\end{equation}
This obeys the PWE component wise, and the transversality condition:
\begin{subequations}
\begin{eqnarray}
&& i\dfrac{\partial}{\partial z} \tilde{\bf A}^{(+)}({\bf x}_\perp,z;\omega)= (-k-\dfrac{\lambdabar}{2}\mbox{\boldmath$\nabla$}_\perp^2)~\tilde{\bf A}^{(+)}({\bf x}_\perp, z;\omega);\nonumber\\ \\\label{vectorPWEa}
&& \mbox{\boldmath$\nabla$}_\perp\cdot  \tilde{\bf A}^{(+)}_\perp({\bf x}_\perp, z;\omega)+\dfrac{\partial}{\partial z} \tilde{\bf A}^{(+)}_3({\bf x}_\perp, z;\omega)\simeq 0.\nonumber\\\label{vectorPWEb}
\end{eqnarray}\label{vectorPWE}
\end{subequations}
In the PWE there is an extra term compared to eq.~(\ref{PWE}) as the factor $e^{ikz}$ has been retained; this in turn is because we wish to avoid extra 
terms in the transversality condition. Moreover this condition can be obeyed only approximately, to the same degree of accuracy as the paraxial 
condition.
The subtleties involved in imposing transversality in the
paraxial regime have been discussed in
reference~\cite{deutsch-pra-91}
This just reflects the fact that eq.~(\ref{long_component}) is an approximate statement.
\subsection*{Gaussian type examples}
To illustrate these ideas further, we consider the case where ${\bf v}({\bf k})$ is proportional to a centred Gaussian factor characterized by a 
width parameter $w$ as in the scalar Laguerre--Gaussian case (\ref{k_LG}):
\begin{equation}
{\bf v}({\bf k})=\begin{pmatrix} {\bf a}_\perp({\bf k}_\perp) \\ c({\bf k}_\perp)\end{pmatrix}e^{-w^2{\bf k}_\perp^2/4},\label{Gaussian_case}
\end{equation}
with ${\bf a}_\perp({\bf k}_\perp), c({\bf k}_\perp)$ polynomial in ${\bf k}_\perp$. Then the transversality condition (\ref{vectorPWEb}) reads
\begin{eqnarray}
&& {\bf k}_\perp\cdot {\bf a}_\perp({\bf k}_\perp)+(k-\dfrac{\lambdabar}{2}{\bf k}_\perp^2)c({\bf k}_\perp)\approx 0,\nonumber\\
{\rm i.e.,}\quad && (1-\frac{1}{2}\dfrac{{\bf k}_\perp^2}{k^2})c({\bf k}_\perp)\approx -\dfrac{{\bf k}_\perp}{k}\cdot {\bf a}_\perp({\bf k}_\perp).\label{transversality}
\end{eqnarray}
The structure of this condition shows again why in the paraxial regime we can impose transversality only approximately. For example, if we imagine 
the polynomials ${\bf a}_\perp({\bf k}_\perp)$ to be preassigned, the expression $(k^2-\frac{1}{2}{\bf k}_\perp^2)$ will almost certainly not be a 
factor in them, so that $c({\bf k}_\perp)$ cannot be a polynomial if eq.~(\ref{transversality}) is demanded as an exact quality. To leading paraxial
order, then, we have
\begin{equation}
c({\bf k}_\perp)\simeq -(1+\frac{1}{2}\dfrac{{\bf k}_\perp^2}{k^2})\dfrac{{\bf k}_\perp}{k}\cdot {\bf a}_\perp({\bf k}_\perp),\label{Gaussian_longitude}
\end{equation}
and as in eq.~(\ref{long_component}) $c({\bf k}_\perp)$ is one order of magnitude smaller than ${\bf a}_\perp({\bf k}_\perp)$.

The scalar Laguerre--Gauss modes of eq.~(\ref{k_LG}) can be extended to define corresponding vectorial photon modes in two ways, with the two 
structures in eq.~(\ref{J3_eigenfns}). Choose an eigenvalue $m\hbar$ for $\hat{J}_3$, and an index $p=0, 1, 2, \ldots$. Then upto an overall 
numerical constant, we can take  ${\bf v}({\bf k})$ to be (in cylindrical coordinates ${\bf k}_\perp\rightarrow (\rho, \varphi)$):
\begin{eqnarray}
&&{\bf v}_{m,p}{({\bf k}_\perp)}=e^{i(m-1)\varphi}\begin{pmatrix}a(\rho) \\ia(\rho)\\c(\rho)e^{i\varphi}\end{pmatrix},\nonumber\\
&&a(\rho)=w~\left(\dfrac{iw\rho}{\sqrt{2}}\right)^{|m-1|}L_p^{|m-1|}(w^2\rho^2/2)e^{-w^2\rho^2/4},\nonumber\\
&&c(\rho)\simeq -\left(1+\dfrac{\rho^2}{2k^2}\right)\dfrac{\rho}{k}a(\rho).
\label{firstVector_LG}
\end{eqnarray}
This corresponds to the $\mbox{\boldmath$\alpha$}_m$ structure in eq.~(\ref{J3_eigenfns}). The other possibility, with the structure of 
$\mbox{\boldmath$\beta$}_m$ in eq.~(\ref{J3_eigenfns}), is:\begin{eqnarray}
&&{\bf v}'_{m,p}({\bf k}_\perp)= e^{i(m+1)\varphi}\begin{pmatrix}ia'(\rho) \\a'(\rho)\\c'(\rho)e^{-i\varphi}\end{pmatrix},\nonumber\\
&&a'(\rho)=w~\left(\dfrac{iw\rho}{\sqrt{2}}\right)^{|m+1|}L_p^{|m+1|}(w^2\rho^2/2)e^{-w^2\rho^2/4},\nonumber\\
&& c'(\rho)\simeq -i\left(1+\dfrac{\rho^2}{2k^2}\right)\dfrac{\rho}{k}a'(\rho).\label{secondVector_LG}
\end{eqnarray}
( The primes here do not mean derivatives with respect to the argument)

\subsection*{General $\hat{J}_3$ eigenfunctions in paraxial
regime} Finally we consider the behavior of the general
$\hat{J}_3$ eigenfunction (\ref{general_eigenfn}) in the
paraxial limit. With the condition (\ref{paraxial_region})
this takes the approximate form\begin{eqnarray} &{\bf
v}_m(k, \theta, \varphi) \simeq \dfrac{1}{\sqrt{2}}\cdot
e^{i(m-1)\varphi}~a(k,
\theta)\left(\begin{array}{c}1-\dfrac{\theta^2}{2}e^{i\varphi}{\cos
\varphi}\\[6pt]i-\dfrac{\theta^2}{2}e^{i\varphi}{\sin\varphi}\\[6pt]-\theta
e^{i\varphi}\end{array}\right)\nonumber\\
&+\dfrac{i}{\sqrt{2}}\cdot
e^{i(m+1)\varphi}~b(k,\theta)\left(\begin{array}{c}1-\dfrac{\theta^2}{2}e^{-i\varphi}{\cos
\varphi}\\[6pt]
-i-\dfrac{\theta^2}{2}e^{-i\varphi}{\sin\varphi}\\[6pt]-\theta
e^{-i\varphi}\end{array}\right),\nonumber\\\label{paraxial_eigenfn}
\end{eqnarray} consistent with eq.~(\ref{long_component}).
From eq.~(\ref{CP2}) we also have, in this limit,
\begin{equation} \hat{\text{S}}_3\
\mbox{\boldmath$\epsilon$}^{(\pm)}(\hat{\bf k})\simeq~
\pm\hbar(1-\theta^2/2)
\mbox{\boldmath$\epsilon$}^{(\pm)}(\hat{\bf
k}),\label{spin_values} \end{equation} for the two terms in
eq.~(\ref{paraxial_eigenfn}). (However, as noted after eq.~(\ref{CP2}), these terms are not eigenfunctions of $\hat{\bf
k}\cdot\hat{\bf S}$ in ${\cal H}_1$).

We see that to order $\theta$ (but not to order $\theta^2$)
the $a(k,\theta)$ term ($b(k,\theta)$ term) in eq.~(\ref{paraxial_eigenfn}) has `spin' angular momentum along
the $z$-axis of amount $\hbar(-\hbar)$, (right/left circular
polarization), so it has `orbital' angular momentum along
the $z$-axis of amount $(m-1)\hbar~((m+1)\hbar)$. But these
are approximate statements. Apart from being valid only to
order $\theta$, there are no consistent definitions of
`orbital' and `spin' \textit{angular momenta} for classical
light or for photons.

\section{Summary and Conclusions}
In this work we have tried to give a comprehensive
treatment, starting from first principles, of the angular
momentum of electromagnetic radiation in both classical and
quantum domains, and covering both general and paraxial
situations.  The free Maxwell field equations are invariant
under the action of the ten parameter Poincar\'{e} group
which leads to the ten basic conservation laws or
COM's. We have developed several ways of expressing them
and in particular in terms of   analytic signals
making the transition to the quantum theory
very easy.

The use of the classical Hilbert space $\mathcal{M}$ 
eq.~$(\ref{setM})$, made up of suitable solutions of the
classical Maxwell equations, serving in the quantum case as
the set of all single photon wave functions 
is a salient feature of our treatment.
For single photons every
component $\hat{J}_j$ of the total angular momentum is in
principle a physical observable, not just the component
$\hat{\bf p}\cdot \hat{\bf J}$ along the momentum direction
as is sometimes stated.  This is made particularly clear by
the properties of the basis of total angular momentum
eigenfunctions $\{{\bf Y}^{(a)}_{lm}(\hat{{\bf k}})\}$ for
photons, to be contrasted with the plane wave energy
momentum eigenfunctions.

The particular focus here has been on the extensively
discussed separation of the total angular momentum of light
into `orbital' and `spin' parts, reminiscent in the
definition of the case of massive particles. We have paid
special attention to their definitions as dynamical
variables, and in the quantum theory to the unitary
transformations they generate on single photon wave
functions. As has been recognized for some time, neither of
them has the algebraic properties -- CR's -- characteristic
of angular momentum in quantum mechanics. This in turn is
related to the well-known result of Newton and Wigner that
the concept of position with reasonable properties is
undefined for photons -- if it were, then so would orbital
angular momentum in the sense of quantum mechanics, and then
spin. In actual fact, the total angular momentum and the
proposed `spin' vector together realize the Lie algebra of
the Euclidean group E(3), though all of them have dimensions
of action. From the theory of the UIR's of this group it is
well known that the eigenvectors of the `translation
generators' $\hat{{\bf S}}$ are `ideal vectors' obeying
Dirac delta function normalization over the sphere
$\mathbb{S}^2$. It follows that normalisable eigenvectors of
$\hat{J}_3$, constructed explicitly in Section V, are never
expressible as discrete linear combinations of $\hat{S}_3$
`eigenvectors'. As can be easily verified, in any single
photon state ${\bf v}_m(k,\theta,\varphi)$ in
eq.~(\ref{general_eigenfn}), $\hat{S}_3$ has an expectation
value in the interval $(-\hbar, \hbar)$ and a \textit{non
vanishing} variance or spread $(\Delta \hat{S}_3)^2$.

The transversality of the free Maxwell field, both of the
field strengths and of the vector potential in the radiation
gauge, is a requirement that has a profound influence on the
quantization process as well as on the description of single
photon states. We have kept careful track of this in the
process of quantization and in the definition and passage to
the paraxial regime. It is to be appreciated that any choice
of a transverse polarization basis
$\{\mbox{\boldmath$\epsilon$}_\alpha({\bf k})\}$ in wave
vector space brings with it a certain degree of phase
freedoms or phase ambiguities, and to this degree the action
of the total angular momentum operator $\hat{J}_3$ on single
photon wave functions will also be affected. For this reason
we have avoided as far as possible any choice of such bases.
We have examined how and in what
sequence these two processes should be implemented,
maintaining consistency from both physical and mathematical
points of view. We have tried to present convincing
arguments to show that in the quantum case the paraxial
limit should be taken after, not before, quantization.

Our treatment of the paraxial limit has been to the lowest
nontrivial order. We have emphasized that most solutions of
the paraxial wave equation are in fact not paraxial; to be
paraxial definite restrictions have to be imposed on a
general solution. This as we have seen has consequences
described in the previous paragraph. Even to this order, we
have seen that the transversality condition makes physical
sense only as an approximate, and not as an exact,
condition. In this way, an appropriate extension of the much
used Laguerre--Gaussian modes from the scalar to the vector
case has been derived.

From the point of view of special relativity, the
appropriate framework for setting up a systematic procedure
to handle higher order paraxial approximations is the front
form of relativistic dynamics.  This is one of three forms
elaborated by Dirac long ago~\cite{dirac-rmp-49}. This
method has been used in the past to handle the polarization
of light consistently to leading paraxial order. A treatment
of higher orders by this method, both classically and in
quantum theory, will be presented elsewhere.

In this work we have emphasized the description and
properties of single photon states, and exploited the fact
that the manifold of solutions of the classical Maxwell
equations goes over naturally to the space of single photon
wave functions. The treatment of two or more multiphoton
states will then follow by using standard quantum mechanical
methods subject to the requirements of Bose statistics. This
and other related aspects will be taken up elsewhere. 
It is our hope that the formalism developed here will
be found useful by those working in the field and in
particular to understand the properties of interesting
quantum situations involving non-classical and entangled 
light with OAM. With our improved understanding of the
algebraic properties of $\hat{\bf L}$  and $\hat{S}$
 in particular that they are
not angular momenta, it seems preferable to expand the OAM
to `Optical Angular Momentum'. 
\section*{Acknowledgements} 
Arvind acknowledges funding from DST India under Grant No.
EMR/2014/000297.  NM thanks the Indian National Science
Academy for enabling this work through the INSA
Distinguished Professorship.

\end{document}